\documentclass{aa}
\usepackage{graphicx}
\usepackage{txfonts}
\usepackage{xcolor}
\usepackage{hyperref}

\hypersetup{
    colorlinks=true,
    linkcolor=blue,    % Color for internal links (e.g., table of contents)
    citecolor=blue,    % Color for citation links
    urlcolor=blue      % Color for external links (e.g., URLs)
}

\usepackage{rotating} % Paket für das Drehen von Objekten

\usepackage{appendix}

\newcommand{\eROSITA}{{eROSITA}}
\newcommand{\SRG}{{\it SRG}\xspace}
\newcommand{\xmm}{{\it XMM-Newton}\xspace}
\newcommand{\chandra}{{\it Chandra}\xspace}
\newcommand{\swift}{{\it Swift}\xspace}
\newcommand{\ROSAT}{{\it ROSAT}\xspace}
\newcommand{\cts}{cts\,s$^{-1}$\xspace}

\defcitealias{2022Buchner}{B2022}

\usepackage{subcaption}         % necessary for continued figures, example in section 3
\usepackage{float}

\begin{document}

\title{The eROSITA DR1 variability catalogue}

\titlerunning{eRASS1 variability}

   \authorrunning{Th. Boller, et al.}

   \author{Th. Boller\inst{1}\thanks{E-mail: bol@mpe.mpg.de}, M. Salvato\inst{1},  J. Buchner\inst{1}, 
   M. Freyberg\inst{1},
   F. Haberl\inst{1}, C. Maitra\inst{1},  A. Schwope\inst{3}, J. Robrade\inst{2}, S. Rukdee\inst{1}, A. Rau\inst{1}, I. Grotova\inst{1}, S. Waddell\inst{1}, Q. Ni\inst{1},  M. Krumpe\inst{3}, A. Georgakakis\inst{4}, A. Merloni\inst{1}, and K. Nandra\inst{1}
          }

   \institute{Max-Planck-Institut f\"ur extraterrestrische Physik, Gie{\ss}enbachstra{\ss}e 1, 85748 Garching, Germany
       \and
   Hamburger Sternwarte, Universit\"at Hamburg, Gojenbergsweg 112,  21029 Hamburg, Germany 
       \and
   Leibniz-Institut für Astrophysik Potsdam (AIP), An der  Sternwarte 16, 14482 Potsdam, Germany
       \and
   National Observatory of Athens, I. Metaxa \& V. Pavlou, P. Penteli, 15236, Athens, Greece
             }

   \date{Received January 26, 2024; accepted June 11, 2025 }

  \abstract
 {

\noindent
With its first All-Sky Survey (eRASS1), the extended ROentgen Survey with an Imaging Telescope Array (eROSITA) on board the Spectrum-Roentgen-Gamma (SRG) mission has offered an unprecedented, comprehensive view of the variable X-ray sky. 
Featuring enhanced sensitivity, broader energy coverage, and improved resolution compared to prior surveys, the eRASS1 Data Release 1 (DR1) catalogue underwent a variability analysis, and in this paper, we performed an advanced variability analysis focusing on a substantial subset of 128,669 sources, all exhibiting a net count exceeding ten.
We performed multiple variability tests, utilising conventional normalised excess variance (NEV), maximum amplitude variability (AMP), and Bayesian excess variance methods (bexvar). The analysis focused on binned light curves; specifically, employing one eroday (a great circle scan with a duration of 4 hours) binning of the German part of the  first eROSITA all-sky survey (eROSITA-DE) data, i.e., the source sample covers only half of the sky.

Within the 128,669 DR1 sources with light curves, our research pinpointed 808 light curves that show hints of variability according to the AMP test, and 298 according to the NEV test. However, after applying suitable thresholds, 90 (123) sources were found to be significantly variable according to the AMP (NEV) tests. In addition, 
1342 sources are considered variable according to the Bayesian test bexvar. The total number of unique sources is 1709, and they form the catalogue of variable sources released with this paper.

We cross-matched with existing X-ray catalogues and identified 258, 318, 598, and 120 sources in 4XMM DR13, 2SXPS, 2RXS, and CSC2.1, respectively. Only 27 sources overlap across all catalogues, while 882 are new X-ray detections from eROSITA DR1. About 70\% are coronal stars, 5\% are Quasi-Stellar Objects, and 1.6\% are normal galaxies. We further subclassified 18 
sources as LMXBs, 11 as HMXBs, and 14 as bright stars.

In this paper, we analyse the variability of eRASS1 sources on a timescale of only a few days. To study the physics of variable sources, we need more deeply pointed observations with other X-ray missions or at least the final depth of the eRASS:8 observations. The timescale of the eRASS1 observations is not representative of the timescales of the expected upcoming eRASS catalogues.

A substantial 52\% of the eRASS1 variable sources were first discovered with eROSITA. The DR1 variability catalogue is excellent for follow-up observations with telescopes such as \xmm, \chandra, or \swift.
}


\maketitle
%
%-------------------------------------------------------------------
\section{Introduction}
X-ray variability serves as a powerful diagnostic tool in astrophysics. It helps identify compact objects, probe extreme physics near black holes and neutron stars, study accretion processes, and reveal explosive events in the universe. These variations give us insight into the cosmos' most energetic and enigmatic phenomena. Rapid X-ray variability often points to compact sources such as neutron stars, black holes, or white dwarfs. For example, the timescale of variability can help estimate the size of the emitting region.
 Also, variability is often associated with material spiralling into a black hole or neutron star via an accretion disc. The timescales of X-ray changes can indicate processes such as turbulence, magnetic reconnections, or changes in the accretion rate.\\
Highly variable X-ray sources in the \xmm slew survey have been analysed by \citet{2022Dongyue}. 
Comprising 265 sources (2.5\%) selected from the XMMSL2 clean catalogue, displayed X-ray variability of a factor of more than ten in 0.2$-$2 keV compared to the second \ROSAT All-Sky Survey (RASS) catalogue (2RXS)
\citep{2016Boller}.
\citet{2020Evans} studied X-ray variability in the Swift/2SXPS\footnote{\url{https://heasarc.gsfc.nasa.gov/W3Browse/swift/swift2sxps.html}} catalogue, which
contains over 80000 variable sources in at least one band or hardness ratio. 
A long-term study of AGN X-ray variability conducted by constructing the structure-function analysis on a \ROSAT-\xmm quasar sample has been compiled by \citet{2017Middei}.
The ensemble X-ray variability of optically selected Quasi-Stellar Objects (QSOs) and the dependence on black hole mass and Eddington ratio has been analysed by \citet{2024Georgakakis}. 
A review of recent developments in modelling one of the most promising X-ray variability and spectral models has provided the relativistic reflection model \citep{2016Dauser}.

Stars are known to be a major contributor to the detected variability in X-ray surveys \citep[e.g. for RASS, see][]{2003A&A...403..247F}. Observed stellar variability is predominantly due to magnetic activity, and the most prominent events, so-called flares, are caused by a large energy release from magnetic structures in the corona. These flare events are analogous to those observed on our Sun, but are more energetic and frequent in active stars. A detailed overview of stellar flares, related phenomena, and underlying physics is given in \citet{2004A&ARv..12...71G}.

X-ray binaries are among the most variable X-ray sources. They include systems with either a low-mass companion \citep[LMXBs;][]{2023hxga.book..120B} or a high-mass companion star \citep[HMXBs;][]{2023hxga.book..143F}.
A comprehensive review of accretion flows in X-ray binaries is given in \citet{2007Done}.
An overview of black hole accretion through time variability is given in \citet{2022DeMarco}.

The X-ray spectral regime variability in Cataclysmic binaries (CVs)  has several causes and occurs on various timescales. Long-term changes on timescales of weeks, months, or years are associated with changes of the mass loss rates from the donor star \citep[e.g. ][]{hessman00}. Short-term changes of the X-ray brightness on timescales of hours or shorter are more typically associated with the accretion geometry.

The accretion regions on and around the accreting white dwarfs might be occulted by the white dwarfs themselves, a phenomenon called self-eclipse;  by the accretion stream/disc; or by the donor star (eclipses). Pronounced variability on such short timescales led to the efficient selection and discovery of magnetic CVs from the 
RASS \citep[][]{beuermann+thomas93}. A few CVs were serendipitously discovered in long observations with the \xmm observatory. The unique periodic pattern of changes in brightness immediately pointed to the nature of the X-ray source even without spectroscopic confirmation \citep{vogel+08,ramsay+09,ok+23}. More recently, pronounced brightness changes between individual eRASS or from scan to scan within one given eRASS have led to the discovery of further CVs, the first to be found with the SRG/eROSITA observatory \citep{schwope+22a,schwope+22b}. This property is addressed in the current paper.

The eROSITA telescope \citep{2012Merloni,2021Predehl}, developed by 
Max Planck Institute for Extraterrestrial Physics (MPE) 
and launched in 2019, provides the first complete all-sky survey in the 0.2$-$8 keV band. 
It has superior energy resolution compared to \ROSAT \citep{1982Truemper} and shows similar sensitivity to \xmm in soft X-rays.
The initial eROSITA all-sky survey (eRASS1) delivers an extensive and unparalleled view of variable X-ray sources spanning the entire sky.
This survey surpasses the flux, energy coverage, and resolution of previous X-ray all-sky missions and marks a significant leap forward in survey variability studies.

This study extensively explores the intricate X-ray variability, specifically concentrating on the eRASS1 light curves acquired from the eROSITA mission. This comprehensive analysis of the eRASS1 data enables a detailed examination of the temporal behaviour of X-ray emission in both coronal stellar and accreting compact object systems.
Our study relies on utilising the initial data release from eROSITA (DR1) \citep{Merloni2024}. From this comprehensive catalogue, we carefully selected and analysed a substantial subset of sources with a net count exceeding ten, resulting in a sizable sample of 128,669 sources for our investigation.

To ensure the integrity of our analysis and avoid any potential overlap, we have judiciously excluded the region around the south ecliptic pole from our investigation. This particular region is being studied extensively in separate research efforts, as detailed in the works by 
\citet{Bogensberger2024} and Liu et. al. (in prep).

The paper is structured as follows: In Sect.\,\ref{sec:dataPreparation}, we introduce the observations and how the parameters used to select the sources were derived. Sect.\,\ref{sec:comparisonMissions} presents a comparison with previous X-ray surveys. 
Section\,\ref{sec:assessing} 
describes the methods for assessing and characterising variability.

Section\,\ref{sec:results} introduces the sources that form the eRASS DR1 catalogue of variable sources, the cross-match with other catalogues, and the X-ray counterparts' characterisation. It also describes the light curves of known objects taken as an example of variability. Section\, \ref{sec:summary} concludes the paper.
The data availability is given in Section\,\ref{sec:data}.

\section{Observations and data preparation}\label{sec:dataPreparation}

The eRASS1 event file underwent meticulous processing and cleaning procedures to conduct our analysis, as detailed in the work by \citet{2022Brunner}. This refined event file has extracted light curves for all 128,669 sources with net counts above 10
in the soft (0.6$-$2.3) keV band.

More in detail, the 
eSASS\footnote{https://erosita.mpe.mpg.de/internal/eROdoc/tasks/srctool\_doc.html\#Light-curve\_file(s)} 
\texttt{srctool} tool computes background counts ($B$), source counts ($S$), 
{backratio, (the factor by which the background counts should be scaled to estimate the number of background counts with the source aperture) and determines the net count rates $X_i$ following the equation:

\begin{equation}
    X_i=\frac{S_i-B_i\times backratio}{f_{\mathrm{exp},i}\Delta t},
\label{eq:eq1}    
\end{equation}

where 
$f_{\mathrm{exp}}$ 
is the fractional exposure time and $\Delta t$ 
is the width of the time bin in seconds. For each data point in the light curve, we have a corresponding value $X_i$.

The task \texttt{srctool} extracts at the given position of the detection, and does not refit the position. The extraction of source and background counts is not the same as the detection, as detection performs deblending to compute the number of net counts, while {\texttt{srctool} works with masked regions.

The effective exposure fraction $f_\mathrm{exp}$ is calculated for each time bin and energy band at a specific source count rate to address instrumental factors influencing source photon counts. This accounts for the fraction of the time bin overlapping with input Good Time Intervals and the fraction of the nominal effective collecting area visible to the source. The calculation considers local telescope vignetting, flux loss due to bad pixels, exceeding the instrumental FoV boundary, and the PSF-wing outside the source extraction region.

Count rate errors $X_{ie}$ in the \texttt{srctool} are determined using the equation:
\begin{equation}
    X_{ie} = \frac{\sqrt{\sigma_{Si}^2 + \sigma_{Bi}^2 \times backratio}}{{f_{exp,i}} \times \Delta t},
    \label{eq:eq2}
\end{equation}

where $\sigma_{Si}$ and $\sigma_{Bi}$ are the uncertainties on the counts in the source region and the background region.

In general, the count error is estimated by \texttt{srctool} as
$\sqrt{\mathrm{counts}}$.
When counts are low ($<25$), \texttt{srctool} does not provide error estimates. In such instances, we estimate the count error as $1+\sqrt{\mathrm{counts}+0.75}$ \citep{1986Gehrels}. This estimation is crucial as excluding such time bins from the variability analysis may introduce bias to the results.

\section{The eROSITA survey scan strategy comparison to other X-ray missions}\label{sec:comparisonMissions}
eSASS provides the data with a time bin of $\Delta t=10$ seconds; however, we have rebinned the data per eroday, which corresponds to a great circle scan with a duration of 4 hours, given that the eRASS1 scan rate is 90 degrees per
hour.
As a result, each source moves across the field of view (FoV) in up to 40\,s, with revisits occurring at an average interval of around 14,400\,s. Typically, six consecutive scans are conducted over locations within the survey area and integrating these scans is referred to as one eroday binning.

The RASS \citep{2016Boller} provides a useful comparison in the context of variability due to its higher cadence (96 minutes versus eROSITA's 4-hour interval) and larger field of view (57 arcminute radius compared to eROSITA's 30 arcminutes). This generally results in around 30 visits within two days for RASS and six visits within one day for eRASS. With its higher response below 0.3 keV, RASS could detect more short-term variability, such as 2-hour-long stellar flares, which eROSITA might miss. 

The 4XMM survey has a similar overall time coverage (less than 1.5 days) as RASS or eRASS, but its sky coverage is smaller due to its focus on serendipitous sources.
Detecting periodic variability on timescales under 400 milliseconds is challenging with eROSITA (which has a time resolution of 50 milliseconds), and long-duration periods over tenseconds are also difficult to detect in eRASS scanning surveys. In these shorter and longer time frames, \xmm is generally preferred due to its ability to provide continuous observation, which is often not feasible for low-Earth orbit telescopes such as \ROSAT or \swift, or scanning instruments such as eROSITA.

\section{Methods}\label{sec:assessing}\label{sec:methods}
\label{sec:VarTest}
In the subsequent sections, we examine the variability tests applied to the eROSITA DR1 point sources. We employ three distinct methods for detecting and characterising variability. These methods are the normalised excess variance (NEV) \citep{Edelson1990,Nandra1997,Edelson2002}, maximum amplitude variability (AMP) \citep{2016Boller}, and Bayesian excess variance \citep[][B22 hereinafter]{2022Buchner}.
Detailed insights into these tests can be found in \citetalias{2022Buchner}, where the authors thoroughly investigate their efficacy for the eROSITA/eFEDS survey. In this context, we provide only a concise overview of each methodology.

\subsection{Normalised excess variability}

The NEV, \citep{Nandra1997}, is determined by calculating the difference between the expected variance  derived from the error bars and the observed variance 
of the net source count rate errors $\rm X_{ie}$ (see eq.\ref{eq:eq2} of this paper):

\begin{equation}\label{eq:NEV}
NEV = \frac{<(X_i - \mu)^2> - X_{ie}^2>}{\mu^2},
\end{equation}

Here $\mu$ represents the mean count rate. It is important to note that the NEV estimator lacks an analytical uncertainty. Extensive numerical studies conducted by \citet{Vaughan2003} led to the identification of an empirical formula for the uncertainty in NEV:

\begin{equation}\label{eq:error_NEV}
NEV_e = \sqrt{\frac{2}{N} \left(\frac{<X_{ie}^2>}{\mu^2}\right)^2 + \frac{4}{N} \left(\frac{<X_{ie}^2>}{\mu^2}\right) NEV},
\end{equation} where  $<X_{ie}^2>$ denotes the mean of the square of the net count rate uncertainties.  N represents the number of data points.

The ratio of $\rm NEV$ to $\rm NEV_e$ offers the significance of the normalised excess variance,
\begin{equation}   
NEV_{\sigma} = \frac{NEV}{NEV_e},
\end{equation} 
in terms of standard deviations $\sigma$. 

According to simulations by \citetalias{2022Buchner}, a threshold of $\rm NEV_{\sigma} > 1.7$ is deemed suitable to avoid false positives, ensuring a very low fraction of false positives 
(c.f. Sect.\,\ref{sec:results}).

\subsection{Maximum amplitude variability}\label{subsec:definitions}
The maximum amplitude variability AMP \citep{2016Boller} is quantified as the range between the most extreme values of the count rate, as expressed by the formula:

\begin{equation}\label{eq:amplmax}
AMP = (X_\mathrm{max} - \sigma_\mathrm{max}) -  (X_\mathrm{min} + \sigma_\mathrm{min}).
\end{equation}
Here, $\rm X_\mathrm{max}$ ($\rm X_\mathrm{min}$) represents the maximum (minimum) count rate, and the associated errors are denoted by $\sigma_\mathrm{max}$ ($\sigma_\mathrm{min}$). The uncertainty in this measure is calculated as:

\begin{equation}\label{eq:error_amplmax}
AMP_{e} = \sqrt{\sigma_\mathrm{max}^2 + \sigma_\mathrm{min}^2}.
\end{equation}
The ratio between $\rm AMP$ and $\rm AMP_e$ provides the significance of the maximum amplitude variability, 

\begin{equation}\label{eq:sigma_amplmax}
 AMP_{\sigma} = \frac{AMP}{AMP_e},
\end{equation}
in units of $\rm AMP_e$. According to findings by \citetalias{2022Buchner},  a threshold of $\rm AMP_{\sigma} > 2.6$ ensures a very low fraction of false positives in our analysis (see Sect.\,\ref{sec:results}).

\subsection{Bayesian excess variance}

The assumption of Gaussianity in calculating the NEV becomes unreliable in the low-count rate regime. To address this limitation, \citetalias{2022Buchner} introduced a Bayesian excess variance estimate (bexvar), designed to operate with Poisson counts instead of inferred uncertainties. In source variability, the Bayesian excess variance model adopts a log-normal distribution for the counts and infers its variance $\sigma_\mathrm{bexvar}^2$. The authors determined that applying a threshold of $0.14$~dex to the lower 10\% quantile of $\sigma_\mathrm{bexvar}$ (a parameter called SCATT\_LO) results in fewer than $0.3\%$ false positives.

The Bayesian excess variance assumes that
at any time bin i, the rate  $X_i$ is distributed according to a
log-normal distribution with unknown parameters (see eq. 18 of \citetalias{2022Buchner}, where
priors for $X_i$ and  $\sigma_\mathrm{bexvar}$ have been chosen uninformative as  wide flat priors; (see eqs. 19 and 20 of \citetalias{2022Buchner}).
The Bayesian framework infers the excess variance from the source region total counts, considering the background region counts at each time step and Poisson statistics. The posterior distribution is computed with the nested sampling Monte Carlo algorithm
MLFriends \citep{Buchner2014stats,Buchner2023} using the
UltraNest\footnote{\url{https://johannesbuchner.github.io/UltraNest/}} package \citep{Buchner2023}.

The intrinsic uncertainties are calculated from a log-normal distribution around a baseline count rate. For both, flat, uninformative priors are assumed.
The intrinsic uncertainties around the mean,
are shown as dashed orange lines in Fig. 9 of \citepalias{2022Buchner}, indicating the upper and lower 1$\rm \sigma$ of the estimated log-Gaussian.

\begin{figure*}[h!]
\centering
        \includegraphics[width=6cm]{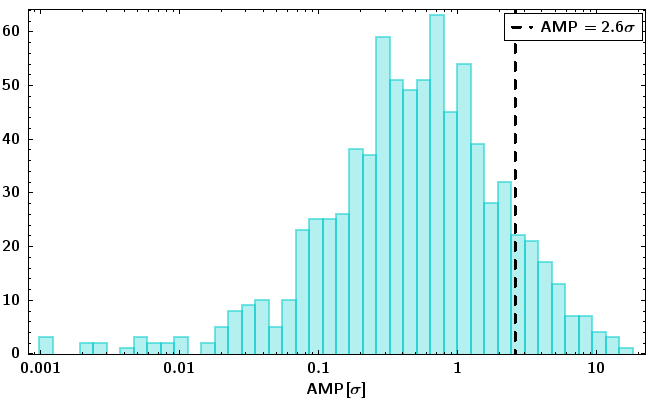}
        \includegraphics[width=6cm]{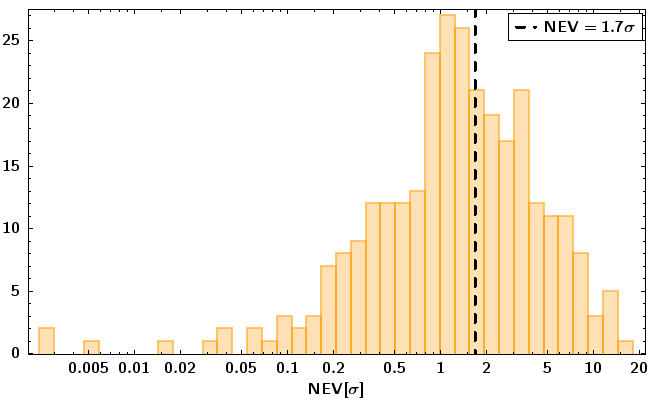}
        \includegraphics[width=6cm]{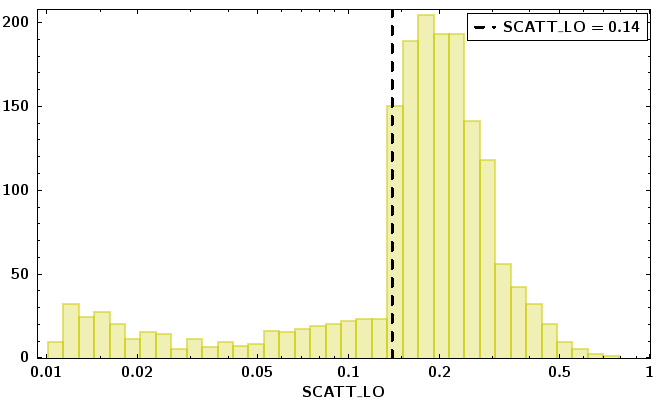}
        \caption{Histograms of the AMP, NEV, and SCATT\_LO distributions in logarithmic scale for the sources released in the catalogue. 
        Negative values are excluded. 
        Vertical lines are thresholds for sources considered significantly variable. More details are given in Sect.\,\ref{sec:samplestats}.
        }
        \label{fig:nev_ampl}
\end{figure*}

\begin{figure}
   \includegraphics[width=\columnwidth]{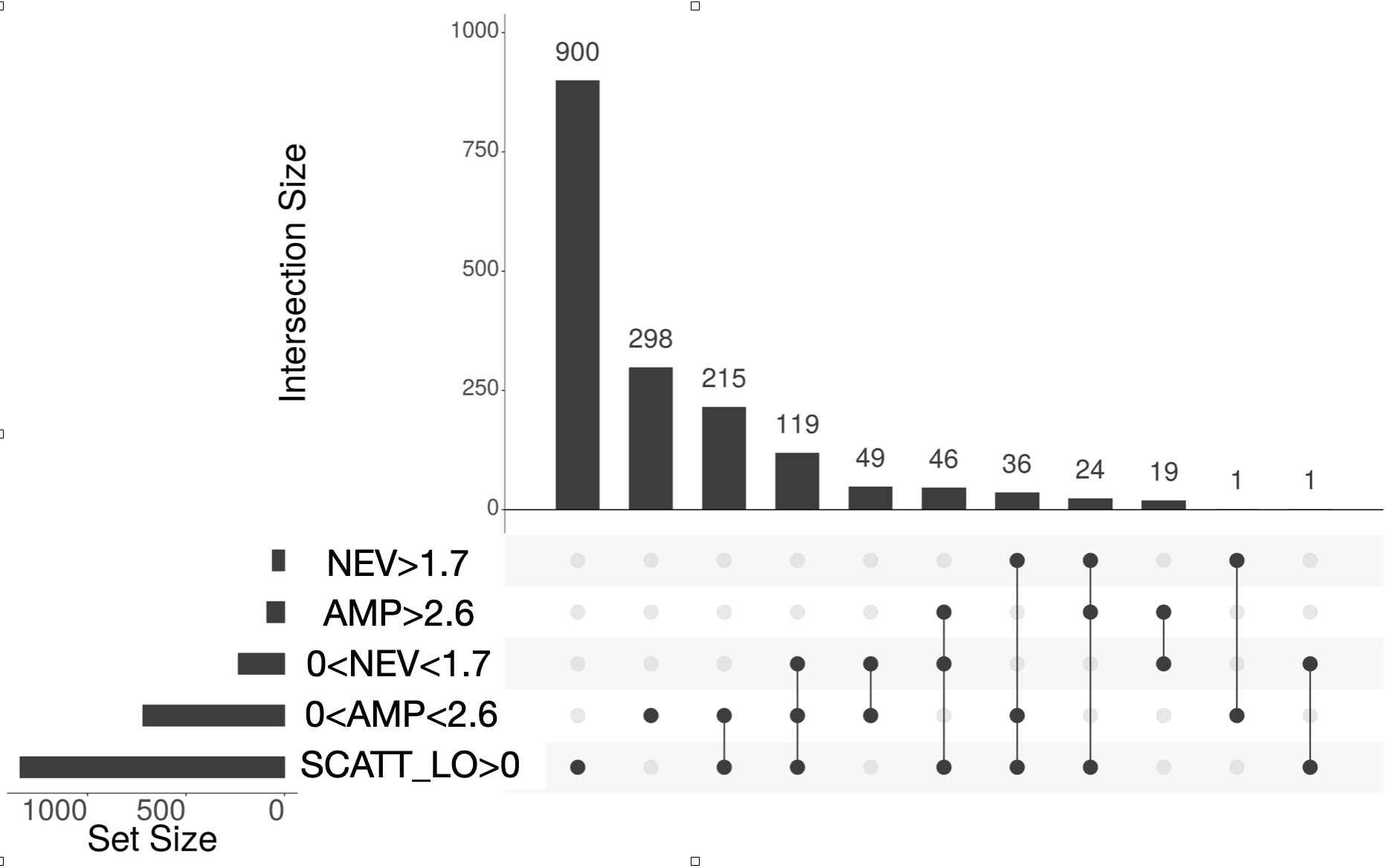}   
  \caption{
  Breakdown of the catalogue of variable sources according to the three tests of variability applied: NEV, AMP, and bexvar. For NEV and AMP, we applied two thresholds: NEV > 0 $\sigma$ and AMP > 0 $\sigma$ indicate sources with some indication of variability; NEV > 1.7 $\sigma$ and AMP > 2.6 $\sigma$ indicate sources that are variable with good confidence. For bexvar we used a single threshold, SCATT\_LO>0.14, that indicates confident variable sources. Out of the 1709 unique sources that comprise the catalogue, only 24 meet all three stringent thresholds. 
  (For more details, see Sect.\,\ref{sec:samplestats}).
  }
\label{fig:pie_diagram}
\end{figure}

\begin{figure*}
        \includegraphics[width=6.2cm]{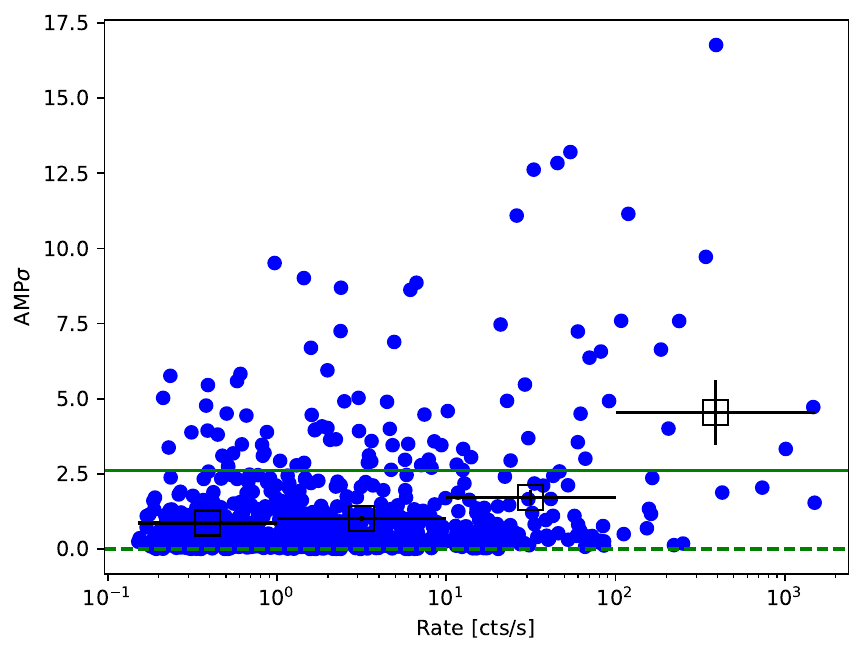}       
        \includegraphics[width=6.2cm]{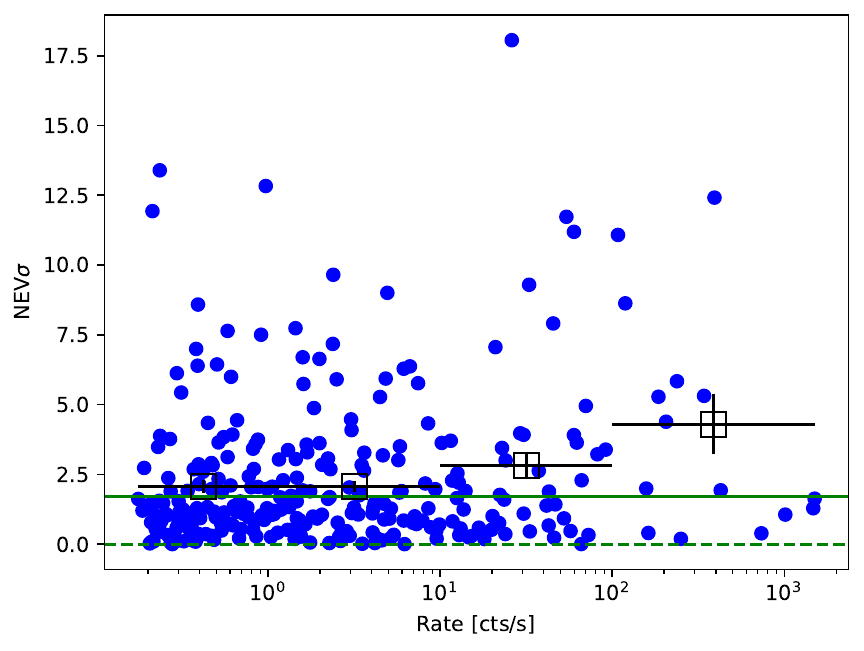}       
        \includegraphics[width=6.2cm]{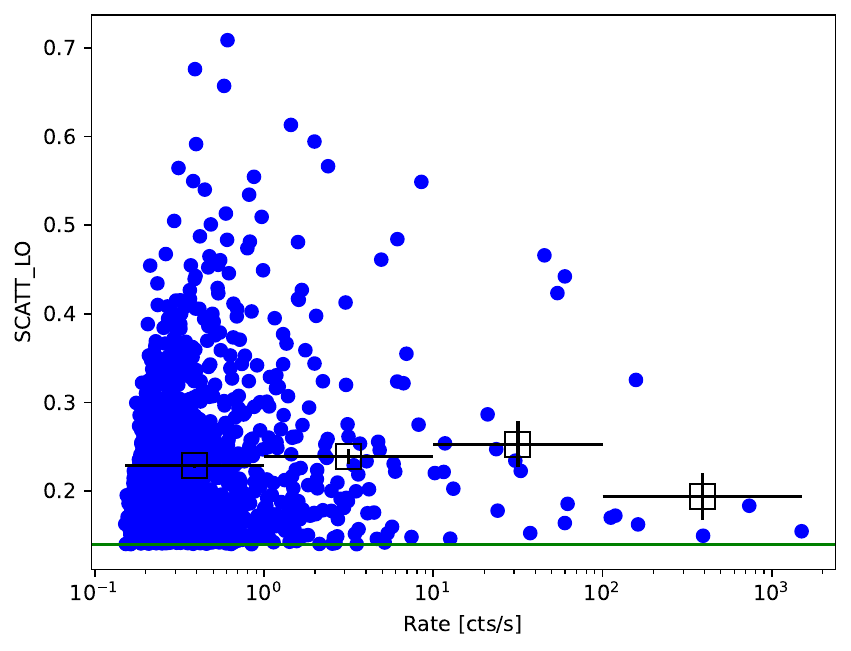}       
      \caption{Distributions of $\rm AMP_{\sigma}$ (left),  $\rm NEV_{\sigma}$ (middle), and SCATT\_LO (right) as a function of count rates. The horizontal solid green lines indicate our significance cuts, and dashed lines indicate preliminary cuts. The black error bars show the mean and standard deviation in four logarithmically spaced bins.
      See more discussion in Sect.\,\ref{sec:samplestats}
      }.
        \label{fig:catalogue_diagnostic_1}
\end{figure*}

\begin{figure}
        \includegraphics[width=9.0cm]{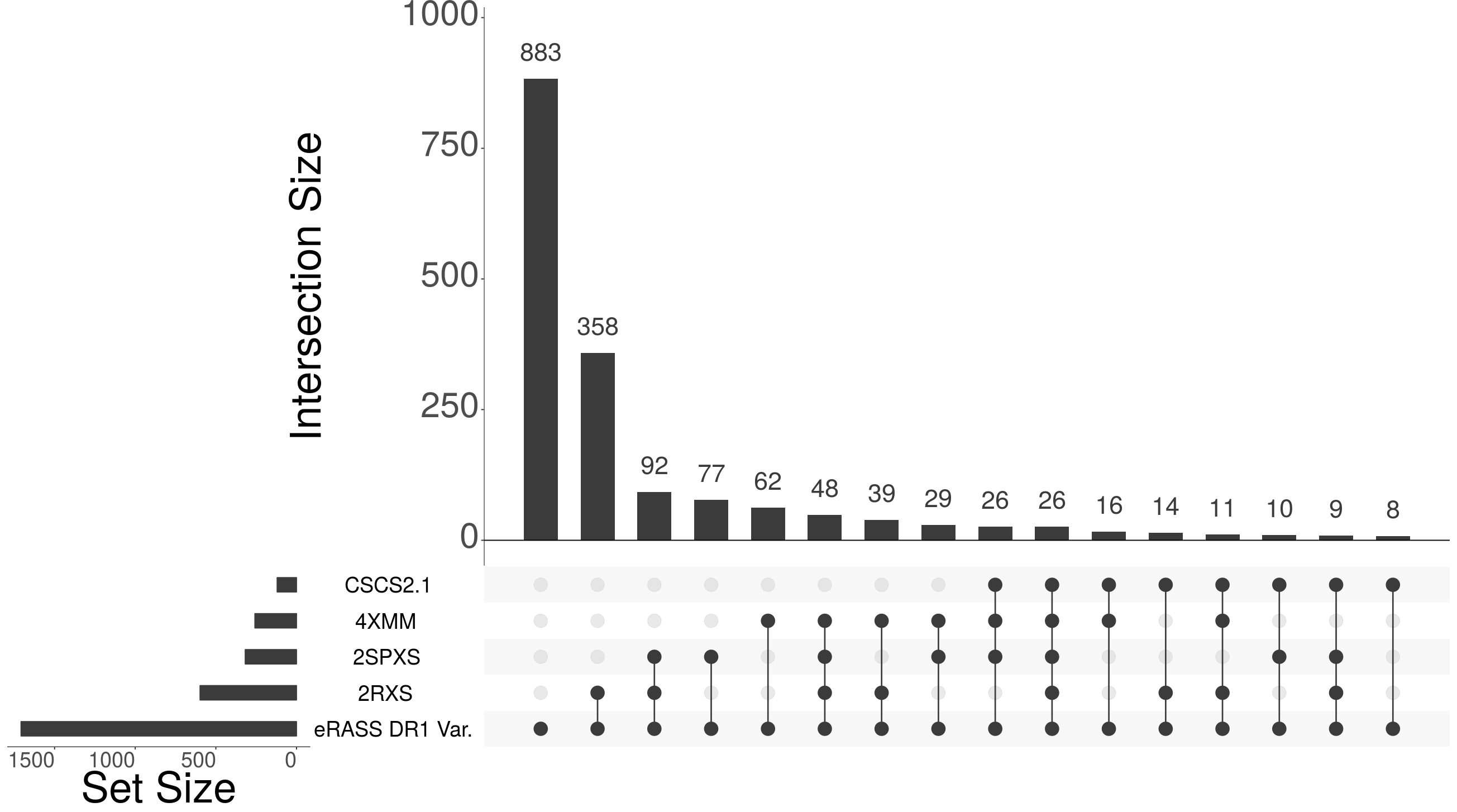}    
        \caption{
 Diagram representing how the 1709 eROSITA DR1 candidate variable sources are distributed among the 4XMM DR13, 2RXS, 2SXPS and CSC2.1 catalogues. 882 (52\% ) of the sources are new detections by eROSITA. The empty intersections are not shown.
            }
        \label{fig:UPSET_surveys}
\end{figure}

\begin{figure}
         \includegraphics[width=\columnwidth]{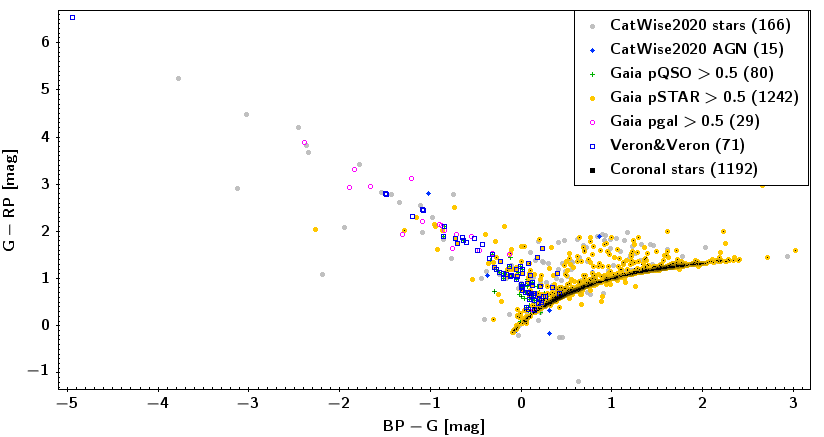}
        \caption{Gaia colour-colour plot of the optical counterparts associated with the eROSITA sources in the  eRASS1 variability catalogue. The classification of the sources as Galactic or extragalactic follows Sect.\,\ref{sec:classification}. Most of the sources are classified as coronal stars from \citet{dr1hamstar}, while for the remaining sources identified in Salvato et al., 2025, the classification is derived 
        from the Gaia properties listed in the original catalogue with a probability of being galaxies, QSOs, or stars higher than 50\%. Few sources are classified as star or AGN based on their WISE colours or as QSO based on the Veron \& Veron catalogue. Most of the sources are stellar, and only 10\% of the sources are classified as of extragalactic nature.}
        \label{fig:Gaia_class}
\end{figure}

\begin{figure*}[!htp]
    \begin{minipage}{\textwidth}
        \includegraphics[width=0.33\columnwidth]{GX_339_4}
        \includegraphics[width=0.33\columnwidth]{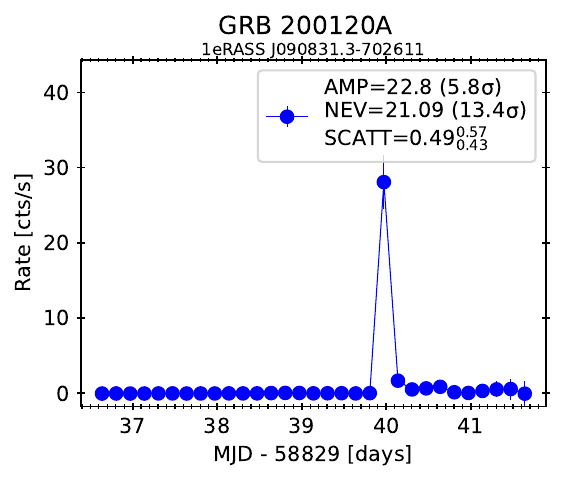}
        \includegraphics[width=0.33\columnwidth]{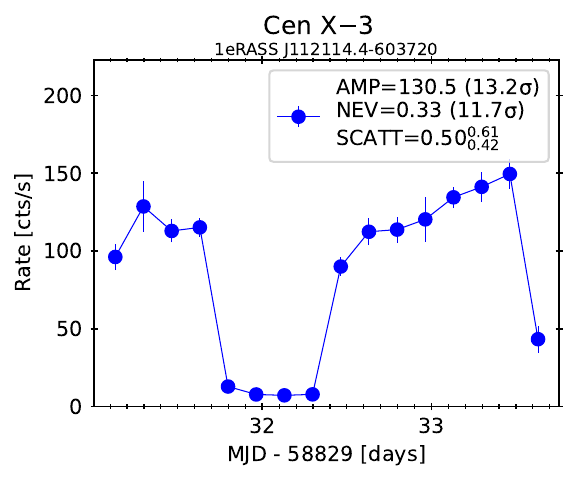}
        \includegraphics[width=0.33\columnwidth]{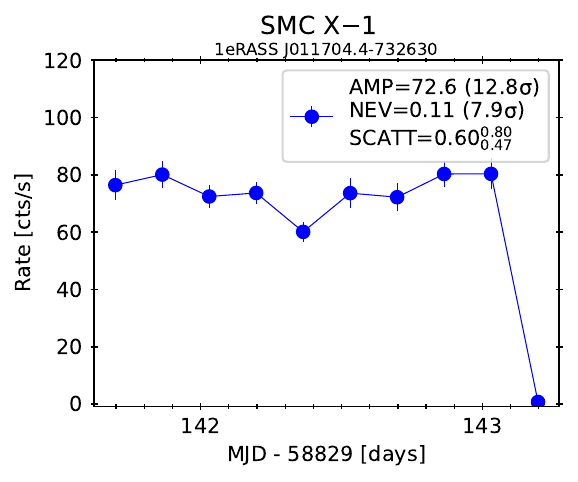}
        \includegraphics[width=0.33\columnwidth]{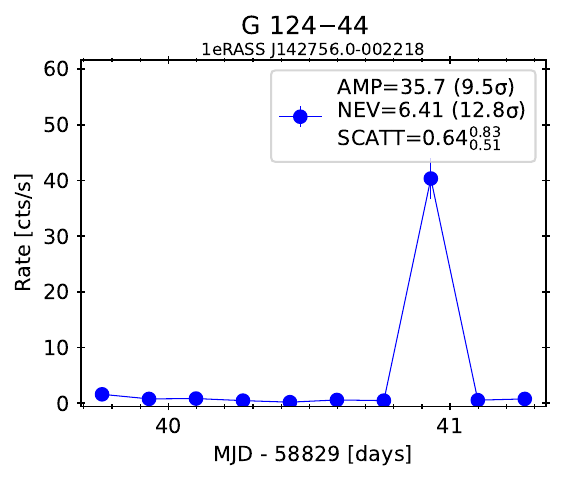}
        \includegraphics[width=0.33\columnwidth]{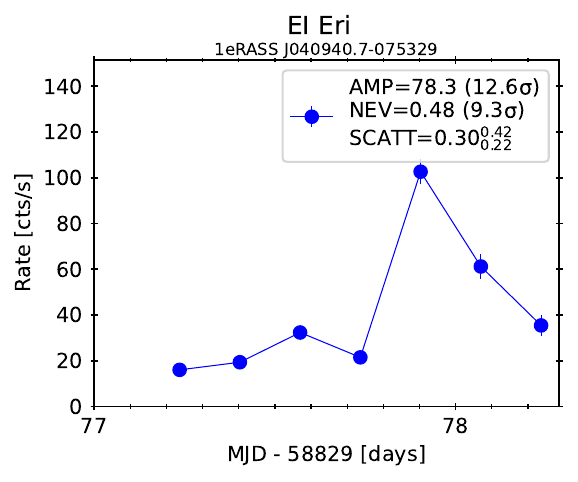}
        \includegraphics[width=0.33\columnwidth]{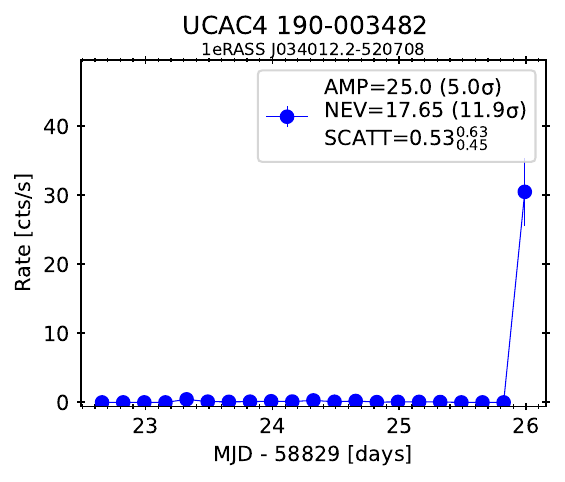}
        \includegraphics[width=0.33\columnwidth]{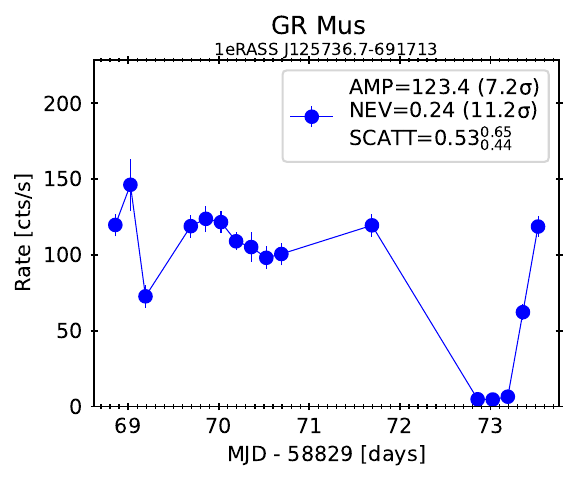}
        \includegraphics[width=0.33\columnwidth]{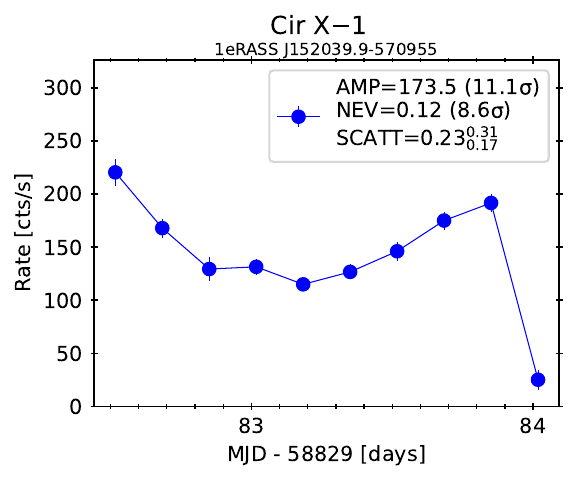}
        \includegraphics[width=0.33\columnwidth]{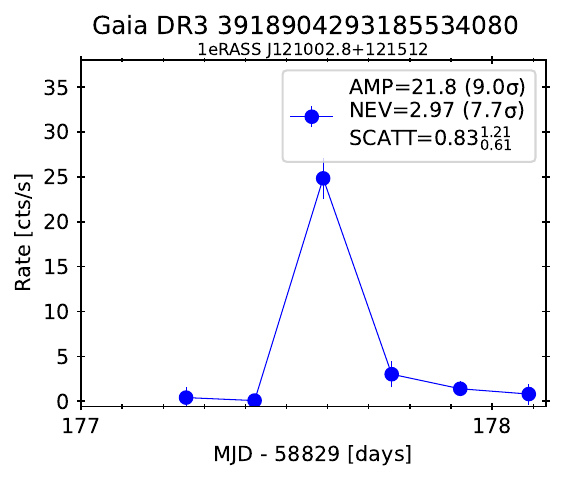}
        \includegraphics[width=0.33\columnwidth]{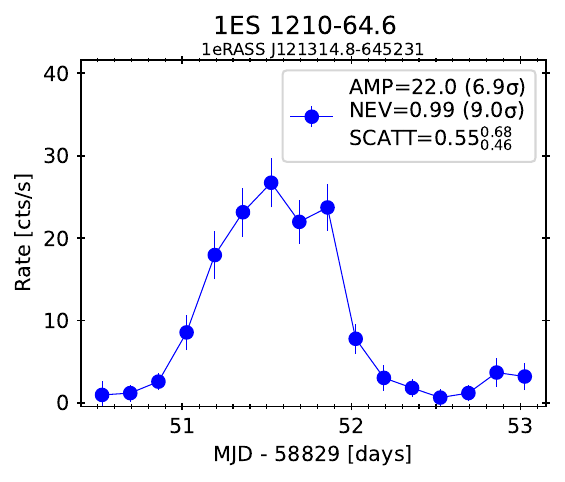}
        \includegraphics[width=0.33\columnwidth]{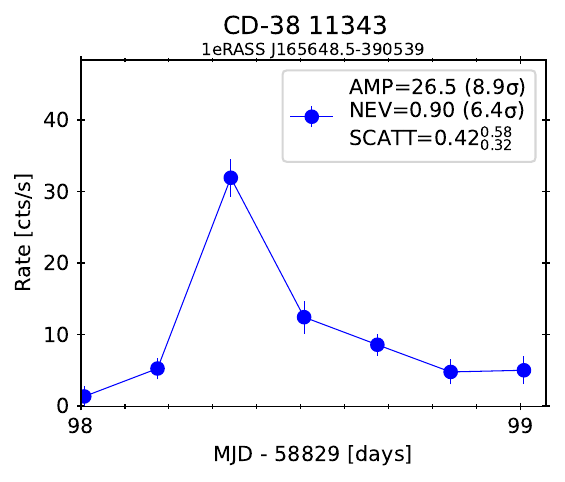}
        \caption{Light curves of the 12 most variable DR1 sources. The bin size is one eroday, and the energy range is (0.6$-$2.3) keV. For the normalised excess variance and the maximum amplitude variability, the values and significance (NEV$_\sigma$, AMP$_\sigma$) in units of $\sigma$ are indicated along the top of each panel. As in Table~\ref{tab:ids_mostvariable}, the objects are ordered by significance (NEV$_\sigma$ or AMP$_\sigma$, whichever is larger) from top left to bottom right.  The Bayesian excess variance is also indicated (SCATT), with the 10\% (bottom, SCATT\_LO) and 90\% quantiles (top).
The IAU name is given at the top of each figure. The x-axis shows the time in MJD units normalised to the launch of eROSITA on 2019-12-11, 21:30 (GMD) in days. The large variability of GX 339$-$4 is interpreted as due to pile-up (c.f. Sect.\,\ref{sec:GX_339}.).}
        \label{fig:mostvariab1}
    \end{minipage}
\end{figure*}

\section{Results}\label{sec:results}

In Sect.\,\ref{sec:samplestats} we describe how the DR1 variability catalogue was built, and this is followed by a section dedicated to the cross-match with existing catalogues of X-ray sources (Sect.\,\ref{sec:crossmatch}), a description of the classification of the optical counterparts (Sect.\,\ref{sec:classification}), and the description of the catalogue data model (Sect.\,\ref{sec:Datamodel}). We conclude with a description of the most variable eRASS1 sources in Sect.\,\ref{sec:brightest}.

At the end of this section, we provide 
notes on cataclysmic variables (Sect.\,\ref{sec:CVs}), notes on the stellar sources in the eRASS1 variability sample (Sect.\,\ref{sect:stars}), notes on X-ray binaries (Sect.\,\ref{sec:binaries}), and notes on AGN in the eRASS1 variability sample (Sect.\,\ref{sec:agns}).

\subsection{The eRASS1 variability catalogue}\label{sec:samplestats}

We applied the three tests to all the 128,669 sources that have at least ten net counts and for which eSASS provided a light curve, over the entire DR1 catalogue, after excluding an area of radius 4.25 degrees, centred on the South Ecliptic Pole (SEP).
This region has the highest number of visits and the deepest data, which needed a dedicated analysis, presented 
in  \citet{Bogensberger2024} and Liu et al. (in prep).

Applying to this sample the criteria described in Sect.\,\ref{sec:methods} we retained:
\begin{itemize}
\item A number of 808 sources with  $\rm AMP_{\sigma}>0$, 90 of which are significant according to the threshold $\rm AMP_{\sigma}>2.6$ of \citetalias{2022Buchner}.
\item A number 298 sources  with $\rm NEV_{\sigma}>0$, 123 of which are significant according to the $\rm NEV_{\sigma}>1.7$ threshold of \citetalias{2022Buchner}.
\item A further 1342 sources found with the Bayesian excess variance test for SCATT\_LO whose values are greater than 0.14 \citetalias{2022Buchner}.
\end{itemize}

The distribution of positive $\rm AMP_{\sigma}$, $\rm NEV_{\sigma}$, and SCATT\_LO values is shown in Fig.~\ref{fig:nev_ampl}. 

Sources from DR1 are considered significantly variable if their $\rm AMP_{\sigma}$ values are greater than 2.6, their $\rm NEV_{\sigma}$ values are greater than 1.7, or their SCATT\_LO values are greater than 0.14 (as indicated by the dashed vertical lines in Fig.~\ref{fig:nev_ampl}.)
These thresholds were derived in \citetalias{2022Buchner} (see his Section 4.1) from enforcing a $3\sigma$-equivalent significance at multiple count rates simultaneously, which resulted in a sample false positive rate that is orders of magnitude lower than 0.3\%. 
Dividing the expected number of false positives in Table 2 of \citetalias{2022Buchner} by their parent sample size of 27910, they estimate a false positive rate of 0.0004\% for AMP and NEV, 0.05\% for the Bayesian excess variance.

Scaling the false positive rates from that study to our parent sample size, we expect fewer than 1 false positives from AMP and NEV and 61 false positives from Bayesian excess variance. 
Sources below these thresholds are candidate variable sources and will be further tested with deeper follow-up X-ray observations.

The total number of unique sources is 1709, and they form the catalogue of variable sources released with this paper. 
The breakdown of the sources among the criteria is shown by using the UPSET\footnote{\url{https://gehlenborglab.shinyapps.io/upsetr/}} tool for the visualisation of intersecting samples \citep{Lex2014}
is shown in  Fig.\,\ref{fig:pie_diagram}.
Only 24  of the 1709 sources are considered variable by the three most stringent criteria simultaneously, while 201 are in common considering the more relaxed definition, and 285 are in common to two tests.

As discussed in \citetalias{2022Buchner}, the different tests are complementary;  for example, the Bayesian excess variance performs best in the low count regime with 900 sources classified as variable only by this method, while the AMP test excels at identifying flares. In this work, five variable sources are identified only through AMP.
Interestingly, the diagram shows that all the sources with NEV>0 and the majority of those with AMP>0, are also classified as varying by at least another method.

Finally, in Fig.~\ref{fig:catalogue_diagnostic_1} we show the dependence of the source count rates on the signiﬁcance of variability according to the tests NEV, AMP, and the Bayesian excess variance, respectively. 
To search for trends, we defined groups in count rates (with bin edges: 0.1, 1, 10, 100, 1000 cts/s). For each group, the mean statistic and its uncertainty are shown by the black error bars in Fig.~\ref{fig:catalogue_diagnostic_1}. 
To test for any significant trend, we test whether these groups share the same mean with an ANOVA test. 
$\rm AMP_{\sigma}$ (p-value < 0.001) and $\rm NEV_{\sigma}$ (p=0.008) show a positive trend, while SCATT\_LO does not. 
This is because the NEV and AMP tests are most sensitive when the error bar sizes are smallest at high count rates, while the Bayesian excess variance method also performs well in the Poisson regime.
The number of variable sources is less than about 1\% of the complete DR1 catalogue.

\subsection{Cross-matches with existing catalogues}\label{sec:crossmatch}

In recent years, many wide-area and all-sky source catalogues have been generated at any wavelength, allowing us to identify and possibly characterise the eROSITA sources in the 
variability catalogue. We have performed cross-matches with various source catalogues following two approaches: one matching eROSITA sources to those in other X-ray catalogues and another matching the most likely multi-wavelength eROSITA counterpart to catalogues of specific sources to characterise their nature.

Concerning the cross-match with existing X-ray catalogues, this allowed us to verify whether or not the eROSITA DR1 variable sources are already known to be X-ray emitters. For that, we matched the eROSITA DR1 variable sources to  \xmm (4XMM\_dr13) \citep{2020Webb}, 2RXS \citep{2016Boller}, 2SXPS \citep{2020Evans}, and \chandra/CSC2.1 \citep[][]{Evans2024}, adopting a radius search of 15" (e.g. about three times the median positional error of eROSITA). We have identified 258, 318, 598, and 79 sources in the 4XMM DR13, 2SXPS, 2RXS, and CSC2.1 catalogues, respectively. Figure\,\ref{fig:UPSET_surveys} shows the sources in common to  the various catalogues using the same tool used in Fig.\,\ref{fig:pie_diagram}.
Only 27 sources are common to all catalogues, and 882 sources (51.5\%) of the eROSITA DR1 variability catalogue are new detections. However, one should not forget that among the tree catalogues used for the comparison, only 2RXS \citep{2016Boller} is an all-sky survey. In addition, the match is done considering only the closest sources to the eROSITA one, and it cannot be ruled out that the match is a chance association, also due to the variable nature of the sources at these wavelengths, and due to the low angular resolution of eROSITA. A more detailed analysis is needed to verify that the association between different X-ray catalogues is correct.

\subsection{Classification of the optical counterparts}\label{sec:classification}

For the characterisation of the sources (star/galaxy classification, which type of star or AGN, etc.), we first relied on the identification of the multi-wavelength counterpart to the entire eROSITA/DR1 catalogue \citep{Merloni2024} presented in \citep{dr1hamstar} and Salvato et al. (in prep.). This shift from eROSITA to the counterparts' optical coordinates significantly reduces the chance of association with existing catalogues of sources or databases such as SIMBAD \citep{Wenger2000}. The drawback is that in the match, it is assumed that the counterpart presented in those papers is the correct one. However, both papers exhaustively discuss the probabilities of associations and  chance probabilities.

In \citet{dr1hamstar}, the authors are hunting for coronal stars and use a  Bayesian algorithm, Hamstar, to determine the most likely coronal star associated with an eROSITA source. The associations are expected to be reliable at 90\% level on average, but increase for sources with high detection likelihood,  as for the sources in the eROSITA DR1 variability catalogue (DET\_LIKE\_0>20) presented here.
Using the \citet{dr1hamstar} catalogue, we found that 1192 (69.7\%) of the eRASS DR1 variable sources are classified as coronal stars. While in \citet{dr1hamstar} the goal was to identify coronal stars, in Salvato et al. (in prep.), the goal was to identify the counterpart of all eROSITA sources, regardless of their Galactic or extragalactic nature. For that, the authors have used NWAY \citep{Salvato2018}, a Bayesian algorithm that assigns the most likely counterpart, accounting for separation between sources in the primary (X-ray) and secondary (optical) catalogues, their positional uncertainties and number densities, and  enhanced by a prior. The prior was defined  using a training sample of X-ray emitters (stars, compact objects, AGN, QSO) with secure counterparts from \xmm and \chandra. Also for this catalogue,  the associations are reliable in more than 90\% of the cases and increase at high detection likelihood.

Using this second catalogue, we find that 1678 (98.2\%, that is 30\% more than with the coronal stars catalogue) of the sources in our sample are reliably associated to a Gaia counterpart. It is reassuring that the 1192 sources in \citet{dr1hamstar} have the same counterparts identified in the Salvato et al. catalogue.
For each source, Gaia provided the probability to be a star (P\_star), a galaxy (P\_Gal) or a QSO (P\_QSO). These probabilities were obtained with a  combination of various machine learning algorithms  applied to mean BP/RP spectra \citep{Creevey2023,Fouesneau2023,Delchambre2023}. We find that 72.7\% of the sources are classified as stars with a probability pStar>50\%, 4.6\% has pQSO>50\%, and 1.6\%  are considered to be galaxies (pGal>50\%). For the latter, the source of the X-ray emission could be from a low luminosity AGN or from ULXs. Both interpretations would explain the X-ray variability. Figure\,\ref{fig:Gaia_class}, shows the distribution of the Gaia counterparts in the BP-G vs G-RP plane, with the breakdown for each class. The arc structure on the bottom right is the locus of the stars, where indeed most of our sources are. The diagonal distribution is the locus of extragalactic sources \citep[see Fig.\,5 of ][]{2019BailerJones}. For completeness, we also cross-matched the counterparts to the Veron \& Veron AGN Catalogue \citep{2010veronveron} for which we found a match for 71 QSO, assuming a maximum separation of 3".
\footnote{Typically, among optical catalogues, the matches are done with a maximum separation of 1". However, most of the sources in the Veron \& Veron catalogue are of nearby extragalactic sources, for which the centre is not extremely precise. For that reason, we opted for a larger radial search.}
 
For the 181 sources, the counterpart is identified in Gaia, but the classification is not provided. For  those, we looked at their Wise W1-W2 catWISE2020 \citep{Marocco2021} colours and found that 166 have colours consistent with those of Galactic stars \citep[$W1-W2\sim 0$;]{Stern2012}. In summary, only 10\% of the 1708 sources forming our catalogue are associated with extragalactic sources. We attempted to further classify the stellar component by looking at the LMXB \citep{2001lmxb}, HMXB \citep{Fortin2023}, Yale Bright Star \citep{1991bsc} and CVs \citep[RKcat;][]{ritter_kolb03} catalogues. More details on the matches to these catalogues of a specific type of sources are presented in Sect.\,\ref{sec:brightest}.

\subsection{eROSITA DR1 variable source catalogue data model}\label{sec:Datamodel}

% shifted before acknowledgement as requested from the editor
%The full catalogue will be available in electronic form at the CDS.\footnote{ftp to cdsarc.u-strasbg.fr (130.79.128.5)} 
%and in this paper, the list of columns and their description is presented in the Appendix in  the Table %\ref{tab:datamodel}.
%
The catalogue includes the unique IAU name, eROSITA source identifier ({\tt DetUID}), 
and the prefix of associated filenames of the DR1 data sets ({\tt filename}) 
along with the eROSITA equatorial, ecliptic and Galactic coordinates.
We provided the standard output values from the eSASS analysis software, for example, the list of source and background net count rates.  
Additionally, we list the variability results for each of the applied  tests
({\tt AMP, NEV, SCATT\_LO}) and the significance of AMP and NEV. The coordinates of the optical counterpart are also provided. 
The key information of the catalogues listed in Sect. \ref{sec:crossmatch}, for example 4XMM\_dr13, 2RXS, 2SXPS and CSC2.1, are also kept.
Finally, the most likely nature of the source (Galactic, extragalactic) is reported, together with the IDs in the original catalogues of LMHB, HMXB, CVs, if available.

\begin{table*}[!htp]
\small
\caption{
%\textcolor{cyan}{
List of the most variable eRASS1 sources, sorted by the larger value between 
$\rm NEV_{\sigma}$ and $\ \rm AMP_{\sigma}$. \\
%\textcolor{red}{replace with IAU names, add a column type}
}
%}
\label{tab:ids_mostvariable}
\begin{tabular}{llrrrrlrl}
\hline
  ID      & IAU name                    & NEV $^a$             & $\rm NEV_{\sigma}$$^b$ &  AMP$^c$         & $\rm AMP_{\sigma}^d$ &SCATT$^f$             & mcr$^g$   & object ID                   \\
          &                             &                     &                        &                  &                      &                      &                           \\
          &                             &                     &                        &                  &                      &                      &                              \\
1$^e$     & 1eRASS J170249.4$-$484724   &  0.10$\pm$0.01      & 12.4                  & 433.2$\pm$ 25.8  & 16.77                & 0.20(0.15, 0.27)       & 588.4     & GX 339$-$4 LMXB              \\
2         & 1eRASS J090831.3$-$702611   & 21.09$\pm$1.57      & 13.4                  &  22.8$\pm$  4.0  &  5.77                & 0.49(0.43, 0.57)       & 1.84      & GRB 200120A                  \\
3         & 1eRASS J112114.4$-$603720   &  0.33$\pm$0.03      & 11.7                  & 130.5$\pm$  9.9  & 13.21                & 0.50(0.42, 0,61)       & 83.73     & Cen\,X$-$3  HMXB              \\
4         & 1eRASS J011704.4$-$732630   &  0.11$\pm$0.01      &  7.9                  &  72.6$\pm$  5.6  & 12.84                & 0.60(0.47, 0.80)       & 80.41     & SMC\,X$-$1  HMXB              \\
5         & 1eRASS J142756.0$-$002218   &  6.41$\pm$0.50      & 12.8                  &  35.7$\pm$  3.7  &  9.52                & 0.64(0.51, 0.83)       & 0.23      & G 124$-$44, Eruptive Variable \\
6         & 1eRASS J040940.7$-$075329   &  0.48$\pm$0.05      &  9.3                  &  78.3$\pm$  6.2  & 12.62                & 0.30(0.22, 0.42)       & 59.55     & EI Eri, RS CVn                   \\
7         & 1eRASS J034012.2$-$520708   & 17.65$\pm$1.48      & 11.9                  &  25.0$\pm$  5.0  &  5.03                & 0.53(0.45, 0.63)       & 0.73      & UCAC4 190-003482, Star        \\
8         & 1eRASS J125736.7$-$691713   &  0.24$\pm$0.02      & 11.2                  & 123.4$\pm$ 17.0  &  7.24                & 0.53(0.44, 0.65)       & 94.04     & GR Mus, LMXB                 \\
9         & 1eRASS J152039.9$-$570955   &  0.12$\pm$0.01     &  8.6                  & 173.5$\pm$ 15.6  & 11.15                & 0.23(0.17, 0.31)       & 162.33    & Cir X$-$1  HMXB              \\
10        & 1eRASS J121002.8+121512     &  2.97$\pm$0.38      &  7.7                  &  21.8$\pm$  2.4  &  9.01                & 0.83(0.61, 1.21)       & 12.52     & Gaia  DR3 3918904293185534080   \\
11        & 1eRASS J121314.8$-$645231   &  0.99$\pm$0.11      &  9.0                  &  22.0$\pm$ 3.2   &  6.89                & 0.55(0.46, 0.68)       & 17.47     & 1ES 1210-64.6, HMXB             \\
12        & 1eRASS J165648.5$-$390539   &  0.90$\pm$0.14      &  6.4                  &  26.5$\pm$ 3.0   &  8.86                & 0.42(0.32, 0.58)       & 22.91     & CD-38 11343, Eruptive Variable \\
\hline
\end{tabular}
\\
Notes:\\    
$^a$ Normalised excess variance and corresponding error ($\rm NEV_e$) \\
$^b$ Normalised excess variability in units of $\rm NEV_e$  \\
$^c$ Maximum amplitude variability and corresponding error ($\rm AMP_e$) \\
$^d$ Maximum amplitude variability in units of $\rm AMP_e$  \\
%$^e$ shortcut of the eSASS light curve-filename; em01\_s1\_820\_LightCurve\_s2\_c010.fits \\
$^e$ The variability is probably affected by pile-up (see Sect.~\ref{sec:GX_339}) \\
$^f$ SCATT\_LO value and the 1 $\rm \sigma$ lower and upper values \\
$^g$ mean count rate in units: $\rm counts\ s^{-1}$ \\
\\
\\
\end{table*}

\subsection{The most variable eRASS1 sources}\label{sec:brightest}

In Fig.~\ref{fig:mostvariab1} and Table~ \ref{tab:ids_mostvariable}, we present the 12 most variable sources within eRASS1, 
ranked according to the larger value between $\rm AMP_{\sigma}$ and $\rm NEV_{\sigma}$, including their best known object ID, using the Gaia identification of the sources, as presented in Sect.\,\ref{sec:crossmatch}. Subsequently, we provide a list of variability characteristics, object identifications, and pre-eROSITA X-ray timing properties. 
Ideally, we would like to provide a spectral analysis for all the sources, but for most of them across the various classes, the number of counts is insufficient.
The DR1 catalogue paper \citep{Merloni2024} adopts a standard energy conversion factor based on an AGN with a photon index of 2 under a Galactic column density of $3\times10^{20}\,\mathrm{cm}^{-2}$, which for 0.6$-$2.3\,keV gives $5.253\times10^{11}\,\mathrm{cm}^2/\mathrm{erg}$. This allowed us to convert the count rates shown in this paper to fluxes.

Some of the brightest stellar flares have sufficient counts to characterise the temperature and luminosity of the plasma. When possible, we fit the data with the flare spectroscopy method, described below.

\subsubsection{Flare spectroscopy}
\label{sec:flarespec}

We fit the DR1 X-ray spectrum following the M dwarf flare analysis method\footnote{Analysis code available at \url{https://github.com/SurangkhanaRukdee/BXA-Plasma}} of \citep{Rukdee2024}. 
This is based on the Astrophysical Plasma Emission Code \texttt{APEC} model \citep{Smith2001}. Briefly, instead of a one or two-temperature model, \citep{Rukdee2024} adopted a source model with a log-Gaussian  Differential Emission Measure (DEM) distribution, where the mean temperature $kT_\mathrm{peak}$, standard deviation $\sigma_{\log T}$, peak normalisation and metal abundance are free parameters, with wide prior ranges. Galactic absorption is modelled with TBabs \citep{Wilms2000} using column densities taken from NH3Dtool\footnote{\url{http://astro.uni-tuebingen.de/nh3d/nhtool}} \citep{Doroshenko2024} based on the map of \citep{Edenhofer2024}.
The spectral fitting is performed with Bayesian X-ray Analysis \citep[BXA][]{Buchner2014}, which connects the nested sampling \citep{Skilling2004,Buchner2023} algorithm MLFriends \citep{Buchner2016b,Buchner2023} implemented in the \texttt{UltraNest} package \citep{Buchner2023} with the fitting environment \texttt{CIAO/Sherpa} \citep{Fruscione2006}. The background spectrum is modelled with the PCA-based model of \citep{Simmonds2018}, as previously described in \citep{Liu2022}, and the background normalisation is fitted simultaneously with the source parameters. We assume Poisson (C-Stat) statistics.

\subsubsection{GX 339$-$4}\label{sec:GX_339}
\begin{figure*}
         \includegraphics[width=9.3cm]{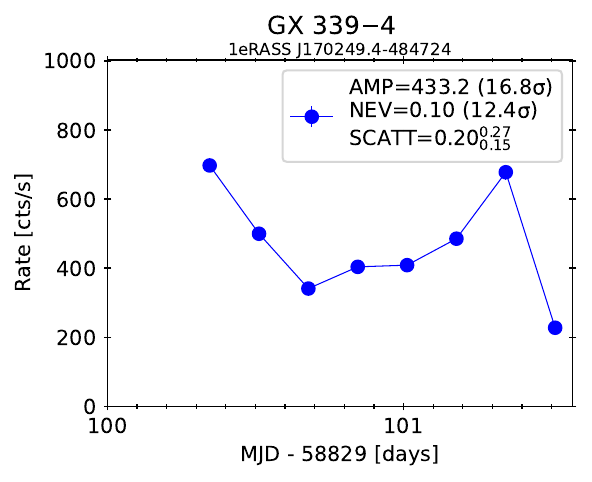}
         \includegraphics[width=9.3cm]{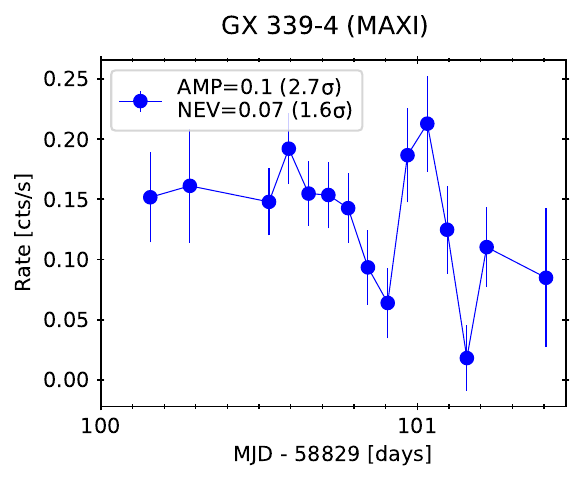}
        \caption{Left panel: eROSITA light curve of GX 339$-$4 in the soft (0.6$-$2.3 keV) band. 
        Right panel:  MAXI light curve in the (2$-$4) keV band. The AMP and NEV values obtained from the MAXI light curve are below three $\rm \sigma$.
        The eROSITA AMP and NEV values are 
        16.77, 12.42 $\rm \sigma$ (soft 0.6$-$2.3 keV) and 3.47, 3.81 $\rm \sigma$ (hard 2.3$-$5 keV), respectively. The eROSITA soft and hard band DR1 light curves are strongly affected by pile-up
        (c.f. Sect.~\ref{sec:GX_339}).
        }
        \label{fig:GX_soft_hard}
\end{figure*}
GX 339$-$4 is the most variable LMXB  (1eRASS J170249.6$-$484725), one of the most well-known and thoroughly investigated LMXBs, at a distance of about 8.4 kpc \citep[e.g.,][]{2016Parker}.
%}
Extensive studies have been carried out to understand state transitions, exploring changes in the shape of the X-ray spectrum, the onset of timing variability patterns, shifts in luminosity, and movements of the inner radius of the accretion disc, as demonstrated by examples such as 
\citet{1972Tananbaum,2010ADroulans}, and \citet{2021Shui}.

Despite the source being classified as variable in eROSITA, the simultaneous light curve from MAXI suggests a non-significant variability as shown in Fig.~\ref{fig:GX_soft_hard}. 
This difference in the light curves may suggest that the variability detected by eROSITA is instrumental and could be due, for example, to the pile-up effect. This effect is generated when two or more photons hit the same CCD pixel in the same (50 ms) read-out cycle. More details about the process related to eROSITA are presented in \citet{Merloni2024}, and a practical example is presented by \citet{2022Koenig} for the observation of  YZ Reticuli, in eRASS2.
Preliminary studies using SIXTE \citep{2019Dauser} indicate that for eRASS1, pile-up starts to become important 
(more than a few percent effect) for point sources brighter than
$\rm 10^{-11}\ erg\ cm^{-2}\ s^{-1}$
in the 0.2$-$2.3 keV band. GX 339-4 has a flux in this band of $\rm 5.48  \times 10^{-10}\ erg\ cm^{-2}\ s^{-1}$.

\subsubsection{GRB 200120A}

The second most variable source is the X-ray afterglow of the gamma-ray burst GRB 200120A detected with Fermi/GBM \citep{2020Hamburg}.
On UTC Jan 20th, 2020, at 23:18:15 (820\,s after the Fermi/GBM trigger) \citet{Weber2021} reported a detection with eROSITA/SRG of
a bright and quickly fading X-ray source. The actual length of the afterglow observations in the eRASS1 all-sky survey is about 40\,s, as indicated in Fig.~\ref{fig:mostvariab1}, 
with an average flux over the period of $1.72\pm0.05\times10^{-12}$\,erg s$^{-1}$ cm$^{-2}$.
The object 1eRASS J090827.6$-$702615 exhibits the highest NEV significance value of all DR1 sources, with $\rm NEV = 13.39\ \sigma$.

\subsubsection{Cen\,X$-$3}
The third most variable source is the HMXB Cen\,X$-$3 (1eRASS J112114.8$-$603723).
Cen\,X$-$3 is the first X-ray pulsar discovered \citep{1971AGiacconi} and has been extensively studied at X-rays to constrain the pulse phase, magnetic field structure, and disc precession \citep{1990Iping,2000ABurderi,2019Ji}.
The light curve is relatively flat with a slightly increasing slope, interrupted by a 50\,ks-long gap where the flux decreased by two orders of magnitude. Another similar decrease is seen at the end of the light curve. The averaged eRASS1 flux is $7.8\pm0.8\times10^{-11}$\,erg s$^{-1}$ cm$^{-2}$.

Cen\,X$-$3 has an orbital period of 2.087\, days and shows eclipses of the X-ray source by the companion star. Extrapolating the ephemeris for the time of mid-eclipse determined by \citet{2023A&A...675A.135K} to the times of the eROSITA observations, we derive MJD 58861.039 days, which is fully consistent with the centre of the observed gap. Therefore, the eclipse of the X-ray source clearly explains the gap and the decrease at the end of the eROSITA light curve about 2 days later.

\subsubsection{SMC\,X$-$1}
One of the first X-ray observations of the HMXB SMC\,X$-$1 have been reported by \citet{1972Schreier}. The source is part of the Small Magellanic Cloud at a distance of about 60 kpc and has been intensively studied with pointed X-ray observations by, for example, \citet{2008Haberl}.

In the first six months of observations, SMC\,X$-$1 (1eRASS J125048.1$-$732630) shows a drop in flux toward the end of the survey due to an eclipse by the B0 supergiant companion. The analysis of the eRASS1,2,3,4 data performed in Sect.\,\ref{sec:binaries} suggests that the extrapolated eclipse times are slightly too early in phase, compared to previous studies.

\subsubsection{G 124$-$44}\label{sec:G124-44}
The source G 124$-$44 is classified as an eruptive variable M4 dwarf type \citep{2015Terrien} at a distance of about 20~pc. During the eRASS1 observations (average flux $7.6\pm0.2\times10^{-12}$\,erg s$^{-1}$ cm$^{-2}$), the object is mostly faint with very low count rates, except for a pronounced single flare reaching a count rate of about 40\,\cts.
To measure the flare parameters, we fit the DR1 spectrum, assuming that the flare dominates the source counts. The spectrum is fitted well by the log-Gaussian DEM model (see Sect.\,\ref{sec:flarespec}) according to a reduced $\rm \chi^2<1$ after rebinning to 50 counts
per bin. We find an abundance of $Z=0.21\pm0.06 Z_\odot$.
The DEM peaks at $3.15\pm0.95$\,keV with a standard deviation of $0.67\pm0.16$\,dex. Such high energies are typical for flares \citep{Robrade2005}. The flare energy is three times higher than that seen in the LTT1445A flare \citep{Rukdee2024} with an identical analysis, possibly related to that flare being 15 times less luminous:
We find a Galactic absorption-corrected flare 0.5$-$2\,keV flux of $5.57\pm0.17 \times10^{-12}$\,erg s$^{-1}$ cm$^{-2}$, which corresponds to a luminosity of 
$\rm L_X=3.16\times 10^{28}$\,erg s$^{-1}$.
The flare is poorly sampled, but if we conservatively assume a flare duration of 40s, we obtain a lower limit of the flare energy of $10^{30}\,\mathrm{erg}$.

\subsubsection{EI Eridani}\label{sec:EIEri}
The object is located at a distance of 54~pc and is classified as a variable star of RS CVn-type binaries and has been extensively studied at X-rays with \xmm pointed observations by \citet{2012Pandey}.  Fig.~\ref{fig:mostvariab1} shows its light curve. The source is already in its pre-flare phase, bright with about 20\,\cts that covers the initial 40~ks, which is followed by an increase up to 100\,\cts and a subsequent decrease in count rate during the remaining eRASS1 coverage. If interpreted as a continuous event, the factor of 5 flare appears to decay over at least 8 hours in the eRASS1 observations.

\subsubsection{UCAC4 190$-$003482}

The star \citep[listed in the UCAC4 catalogue;][]{Zacharias2012}}\footnote{\url{https://www.stellarcatalog.com/stars/ucac4-190-199094}} has a distance of 25.8 pc and is classified as a red dwarf. To our knowledge, it was not previously detected in X-rays. The star shows a strong increase up to 30 $\rm counts\ s^{-1}$ in the last eroday. Before that, the light curve shows very low activity below 2 $\rm counts\ s^{-1}$, indicating a phase of inactivity (see Sect.\,\ref{sect:stars}). We quantify the flare spectroscopically as described in Sect.\,\ref{sec:flarespec}. The abundance is $Z=0.59 \pm 0.22 Z_\odot$. Similar to G~124$-$44, the DEM distribution peaks at $3.14\pm0.98$\,keV with a width of $0.62 \pm 0.20$\,dex. This gives a flux of $0.50\pm0.04 \times10^{-12}$\,erg s$^{-1}$ cm$^{-2}$, which corresponds to a luminosity of $\rm L_X=2.82\times 10^{27}$\,erg s$^{-1}$. The flare is likely not fully sampled by eROSITA. If we conservatively adopt a flare duration of 40\,s, we obtain a lower limit for the flare energy of $10^{29}\,\mathrm{erg}$.

\subsubsection{GR Mus}
GR Mus is an LMXB \citep{2006Qian}. 
The light curve shows erratic behaviour with highly significant increases and decreases. Towards the end of the light curve, three data points with a two orders of magnitude decrease in flux are seen. No data points are available from the immediate scans before, so this behaviour lasted at least two erodays (8\,hours) and recovered to the original flux levels within another 8\,hours.

GR Mus (also known as XB 1254–690) has an estimated distance between 8 and 13 kpc \citep{1987Motch}, and it is known to show dipping behaviour in X-rays, which regularly repeats with a period consistent with the orbital period ($\sim$3.9 hours) of the binary system. However, the dips vary in duration and depth and can disappear completely \citep[see ][and references therein]{2009A&A...493..145D}. A precessing accretion disc has been discussed as the possible origin of the variability \citep[][]{2013Cornelisse}. The orbital period of the LMXB is close to the eROSITA scanning period of 4.0\,hours (resulting in a beat period of $\sim$10\,days; for other examples of beat effects with the regular eROSITA sampling see Sect.\ref{sec:CVs}). Therefore, the three low-flux measurements were taken at a similar orbital phase and do not necessarily indicate a continuous period of low flux. We suggest that the low fluxes were caused by deep dips that repeated over three consecutive binary orbits. We cannot say how many orbits in total the dips were present, because due to the slow phase drift, the high-flux data would then be taken outside dip phases. We conclude that the eROSITA light curve is consistent with the behaviour seen from this source before, with deep dips seen over consecutive binary orbits.
The eRASS1 catalogue lists the average flux over the period shown in Fig.~\ref{fig:mostvariab1} as $8.7\pm1.3\times10^{-11}$\,erg s$^{-1}$ cm$^{-2}$.

\subsubsection{Cir X$-$1} 
The HMXB Cir X$-$1 has been studied with EXOSAT \citep{1986Tennant} to constrain the X-ray outburst characteristics, and physical mechanisms have been investigated by \citet{1972Blumenthal}. The source was bright during the eRASS1 observations, with a pronounced drop during the last eroday. Cir X$-$1 is known to switch between high and low X-ray states \citep[e.g.][]{2012A&A...543A..20D}. The eROSITA light curve may indicate an ingress into such a low state within 4\,hours, but no further constraints can be derived. The eRASS1 catalogue lists the average flux over the period shown in Fig.~\ref{fig:mostvariab1} as $1.5\pm0.1\times10^{-10}$\,erg s$^{-1}$ cm$^{-2}$.

\subsubsection{Gaia DR3 3918904293185534080}

The Gaia counterpart to this X-ray source has a high proper motion\footnote{ \url{https://ui.adsabs.harvard.edu/abs/2020yCat..36490006G/abstract}}, at a distance of 59.74 pc. This is the first X-ray detection of the source. The eROSITA discovery light curve shows a strong flare lasting for one eroday, followed by a decline in the X-ray count rate values towards the end of the observations. This is similar to K/M dwarfs found in eFEDS \citep{2022Boller}. However, Gaia does not provide type information on the source. 
When quantifying the flare spectroscopically, we find an abundance of $Z=0.46 \pm 0.12  Z_\odot$. Similar to the other two stars before, the DEM distribution peaks at $3.81\pm0.74$\,keV with a width of $0.55\pm0.08$\,dex. This gives a flux of $9.33 \pm 0.28\times10^{-12}$\,erg s$^{-1}$ cm$^{-2}$, which corresponds to a luminosity of $\rm L_X=5.37\times 10^{28}$\,erg s$^{-1}$. If we crudely adopt a flare duration of 4h, assuming that the peak was sampled and the subsequent light curve data point in Fig~\ref{fig:mostvariab1} is elevated because it traces the end of the flare, we obtain a flare energy of $7\times10^{32}\,\mathrm{erg}$.

\subsubsection{1ES 1210-64.6}
The source 1ES 1210-64.6 is an HMXB, which has also been detected at
hard X-rays (17$-$60 keV) with INTEGRAL \citep{2010Krivonos}. The object, with a distance of about 1 kpc, has been studied with different X-ray satellites \citep{2009AMasetti,2014Coley}. At the start of the eRASS1 observations, count rates are consistent with zero. Then the source shows an impressive X-ray flaring event lasting over about tenerodays, reaching up to about 25 $\rm counts\ s^{-1}$. At the end of the eRASS1 observations, the rate decreases to a few counts per second. The eRASS1 catalogue lists the average flux over the period shown in Fig.~\ref{fig:mostvariab1} as $1.63\pm0.03\times10^{-11}$\,erg s$^{-1}$ cm$^{-2}$.

\subsubsection{CD$-$38 11343}

CD$-$38 11343 is classified as an eruptive variable star and previously detected in the 2RXS catalogue \citep{2016Boller}. It is an M3Ve + M4Ve binary system of flare stars \citep{2014Tamazian}. The 2RXS light curve does not show any significant variability, and a spectral fit with an optically thin metal emission model gives a temperature of 0.83$\pm$0.10 keV. During the eRASS1 observations one pronounced flare has been detected with a peak count rate of about 30 $\rm counts\ s^{-1}$. The eRASS1 catalogue lists the average flux over the period shown in Fig.~\ref{fig:mostvariab1} as $2.13\pm0.05\times10^{-11}$\,erg s$^{-1}$ cm$^{-2}$.

\subsection{Notes on cataclysmic variables in the eRASS1 variability sample}\label{sec:CVs}
CVs are semi-detached binaries with white dwarfs that accrete matter via Roche-lobe overflow from low-mass, main-sequence stars \citep{warner95}. They were found in reasonably large numbers  by their optical variability through nova and dwarf nova outbursts, by their blue colour in objective prism surveys, spectroscopically in the SDSS as quasar candidates, and as counterparts to point-like X-ray sources \citep{gaensicke05}. Indeed, all CVs have in common that they are X-ray sources. This property allows for the composition of flux-limited samples and will eventually allow for the composition of volume-limited samples to constrain close binary evolution \citep{schwope+02, Schwope+2024}. 
The accretion process in CVs generates intense X-ray emission, characterised by variability over a wide range of timescales. This variability is key to understanding the dynamics of the accretion flow and the interactions between the white dwarf and its environment. We found 16 sources matched to the CVs catalogue of \citet{ritter_kolb03}. The sources can be subclassified as tenmagnetic CVs (5 intermediate polars, 5 polars), 5 dwarf novae (2 SU UMa, 2 U Gem, 1 Z Cam subtype), and one non-magnetic nova-like object of the VY Scl subtype. All but two were formerly detected in 2RXS. The prevalence of the magnetic objects in the small sample is due to their intrinsic short-term variability on the orbital or the spin period of the accreting white dwarfs. 

The observed pattern in the light curves reflects the folding of the intrinsic variability with the sampling period of the spacecraft. Intrinsic variability happens during the orbital period and is caused by eclipses, self-eclipses and perhaps other geometrical effects, plus variations in the instantaneous mass accretion rate, which may give flares or at least non-repetitive brightness at a given spin or orbital phase. The sampling pattern may thus lead to irregular and, in some cases, very regular brightness variations in DR1. 

A nice example of a regular light curve with a simple on-off pattern is that of 
1eRASS J031130.6$-$315250 an AM Herculis star or polar with a binary orbital period of $0\fd05872$. 
The DR1 light curve is shown in Fig.~\ref{fig:Fig_CV_Maitra}.

\begin{figure}
        \includegraphics[width=\columnwidth]{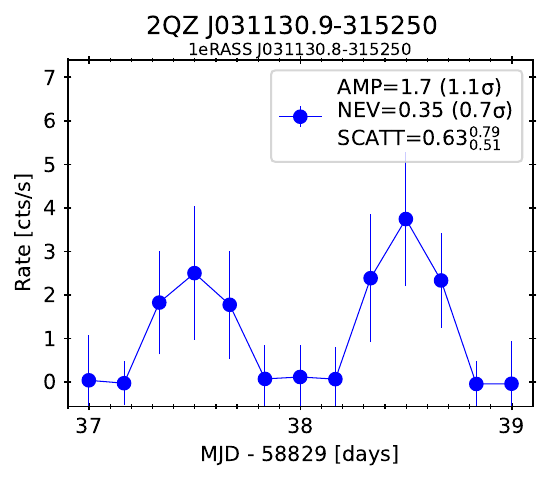}
        \caption{
DR1 light curve of 1eRASS J031130.8-315250 (aka  2QZ J031130.9-315250) as an example of a typical light curve 
with a simple on-off pattern of  an AM Herculis star or polar with a binary orbital period of $0\fd05872$ 
(see Sect.\,\ref{sec:CVs} for details).}
        \label{fig:Fig_CV_Maitra}
\end{figure}

The implied DR1 period of about 24 hours is much longer and likely occurs on some beat frequency between the orbital period of the CV and the scan period of the \SRG spacecraft 
\citep[see][for another striking example of such an apparent long-period variability of an ultra-compact double degenerate binary]{Maitra2024}. While a quantitative variability analysis and period search could be performed, the light curve shows only two humps, making the result highly unreliable. The scanning pattern mentioned is a further complication. Therefore, we defer to follow-up studies to quantify the significance and period.

\subsection{Notes on stellar sources in the eRASS1 variability sample}\label{sect:stars}

Stellar sources are the majority of the observed X-ray variable sources (see Fig.\,\ref{fig:Gaia_class}) and contribute substantially to the number of sources discussed in this work.
Especially coronal emitters that generate X-rays by magnetic activity are known to be highly variable and, as detailed in Sect.\,\ref{sec:crossmatch}, 69\% of the sources presented here are identified as coronal stars in the catalogue of \citet{dr1hamstar}.
The most extreme luminosity and energy among the three stellar flares analysed here are still within the range of flares found in eFEDS \citep{2022Boller}, although the \eROSITA\ time sampling likely prevents capturing the full flare.

Many flaring stars, young active stars, and active binaries such as RSCVn systems are found to be among the most variable sources. Several add to the list of the most variable eRASS sources reported in Table~\ref{tab:ids_mostvariable}. Exemplary light curves of stellar sources are shown in Fig.~\ref{fig:mostvariab1}, and some prominent examples are discussed below. Judging from the 'object type' as given in the SIMBAD database and their Gaia colours, nearly half of the significantly variable stars are M~dwarfs; in addition 10\,--\,20\,\% are classified as very young or pre-main sequence stars and a similar fraction as multiple systems or active binaries. Roughly another 10\,\% belong to other types of variable stars, and some sources are unclassified, apparently ordinary main-sequence stars.

The most variable stellar object is the RSCVn variable V*~EI~Eri (see Sect.\,\ref{sec:EIEri}), while the second most variable stellar object is the eruptive variable star G~124-44 (see Sect.\,\ref{sec:G124-44}). During the eRASS1 observations, the object is mostly faint except for a pronounced single flare.

A similar event is seen on UCAC4~190-003482, which remained inactive throughout the eRASS1 observations except for the last eRASS scan, where it exhibited a sudden brightness increase, reaching a count rate of 30\,\cts.
These burst-like, strong but short flaring events typically occur on timescales of minutes up to an hour and thus mostly produce highly elevated count rates in single eRASS scans (see Fig.~\ref{fig:mostvariab1}). Late-type stars, such as the ones shown here, M dwarfs, are frequently identified as their originating sources.

Flaring and active periods may also be present for longer timescales of hours to days. The star 2MASS J12100301+121507 (Gaia DR3 3918904293185534080), a mid-M dwarf at a distance of 60~pc, shows an extreme flare with rates up to 25\,\cts that lasts for several hours and is still declining at the end of its survey coverage. Another example among the most variable DR1 objects is the nearby M~dwarf binary CD-38~11343 (GJ~2123) at 16~pc distance. It is a bright, variable X-ray emitter as shown by its light curve in Fig.~\ref{fig:mostvariab1}.

These findings on stellar variability are similar to those from the eFEDS variability search of \citet{2022Boller} or those described in a detailed analysis of the eROSITA field-scan on the $\eta$ Chamaeleontis cluster given by \citet{2022Robrade}.
The variability discussed there is likewise associated with flaring due to magnetic activity in late-type stars, which dominate the X-ray stellar variability. Notably also several early-type stars also exhibit distinct X-ray variability. While in 'normal' O/B stars, X-ray emission, which is attributed to wind-shocks, is rather constant, massive stars are diverse and more variable sub-categories exist. However, due to their faintness, lower amplitudes and longer timescales, they are basically absent in this work.
A systematic eROSITA survey of Be stars by \citet{2023Naze} presents several examples of eRASS variability on scan (4~h) as well as survey (0.5~yr) timescales.

\subsection{Example of eRASS1 X-ray variability for X-ray binaries: the case of SMC\,X$-$1}\label{sec:binaries}

The large AMP value obtained for the HMXB SMC\,X$-$1 is caused by a drop in flux in the last data point of the eRASS1 light curve (see Fig.~\ref{fig:mostvariab1}). In SMC\,X$-$1, a neutron star orbits a B0  supergiant, which eclipses the X-ray source regularly every 3.89 days, the binary orbital period of the system \citep{2015A&A...577A.130F}. Using the ephemeris from \citet{2015A&A...577A.130F}, we obtain an orbital phase of -0.0505 (phase 0 corresponds to mid-eclipse time) for the time of the last eRASS1 scan and -0.0933 for the scan before (which is still at the pre-eclipse flux level). 

To investigate this further, we did a detailed analysis of the complete eROSITA data of SMC\,X$-$1 from eRASS1-4. To create light curves and spectra, we extracted events from circular regions around the source and a nearby background region. Because of the source brightness, we chose a sufficiently large circle with 2\arcmin\ radius. The four eRASS spectra were simultaneously fitted by an absorbed two-component model, comprising a power law and diskbb emission. The derived unabsorbed fluxes were used to convert the count rates to X-ray luminosities, assuming a distance of 65\,kpc. Figure~\ref{fig:SMCX1_LX} shows the X-ray luminosity versus time in Modified Julian Date (MJD). Similarly to eRASS1, also the last eROSITA scans of eRASS4 fall near the eclipse ingress of SMC\,X$-$1. \citet{2015A&A...577A.130F} have derived eclipse times and durations for eclipse ingress and egress from hard X-ray observations (17$-$40\,keV), which do not suffer from photo-electric absorption effects.
In Fig.\,\ref{fig:SMCX1_LX} we mark the expected times for the start of eclipse ingress (green dashed lines) and the start of the full eclipse (dash-dotted green lines) using again the ephemeris from \citet{2015A&A...577A.130F}. Their uncertainties are less than 2.6$\times 10^{-4}$\,days and negligible. During eRASS1, the last two data points of the light curve are fully consistent with the eclipse ingress between them. During eRASS4, the last but one point falls into the expected hard X-ray ingress period, which may indicate a start of the full eclipse slightly earlier than predicted by the ephemeris. However, at lower X-ray energies (eROSITA is most sensitive between 0.5 and 2\,keV), the eclipse is less sharp and the low luminosity could also be explained by strong absorption effects often seen at these orbital phases.
Unfortunately, no egress was covered, and given the uncertain eclipse duration in the eROSITA energy band, no further constraints can be derived. 

\begin{figure*}
     \begin{center}
        \includegraphics[width=0.245\hsize]{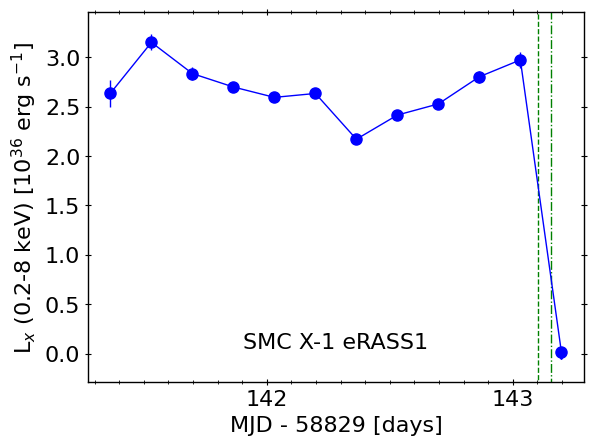}
        \includegraphics[width=0.245\hsize]{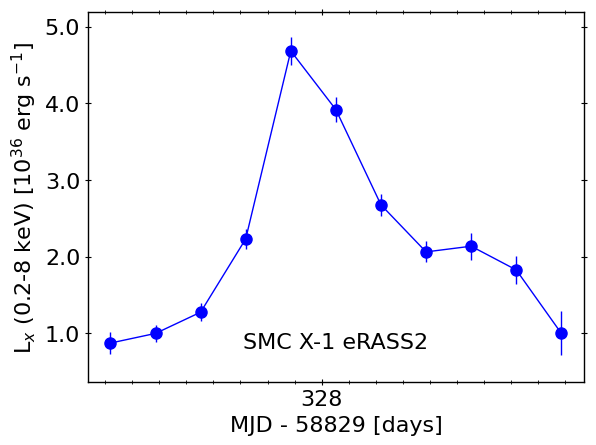}
        \includegraphics[width=0.245\hsize]{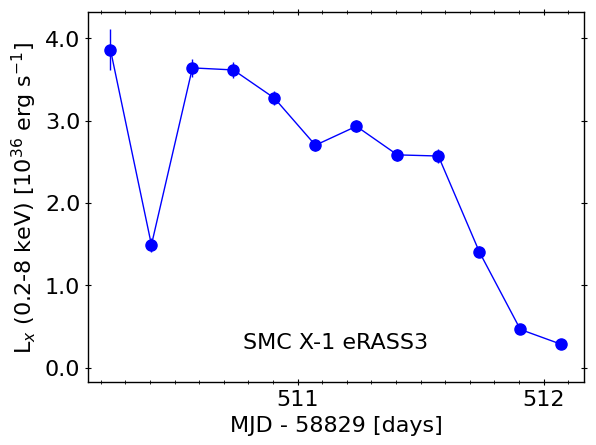}
        \includegraphics[width=0.245\hsize]{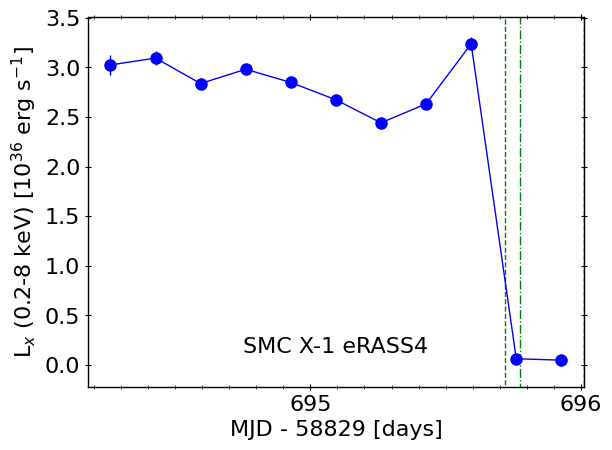}
     \end{center}
        \caption{eROSITA light curves of SMC\,X$-$1. Count rates were converted into X-ray luminosity using parameters from spectral modelling. The vertical dashed and dash-dotted green lines mark the orbital phases of the expected start of eclipse ingress and start of full eclipse, respectively (see Sect.\,\ref{sec:binaries}).
        }
        \label{fig:SMCX1_LX}
\end{figure*}

\subsection{Notes on AGN in the eRASS1 variability sample}\label{sec:agns}
As mentioned, about 11\% of the DR1 variable sources are classified as AGN. 
This is in line with the AGN fraction of variable sources found in the eFEDS survey \citep{2022Boller}, which probed similar time-scales of hours at deeper exposures with the same soft X-ray instrument.
The variability of AGN shows a bending powerlaw power spectrum, in analogy to that of X-ray binaries \cite[e.g.][]{McHardy2006,Ponti2012}.
Various models for short-term AGN variability have been proposed, e.g.
the light bending model \citep{2004lightbending}, the hot spot model \citep{1997hotspots}, or the 
Bardeen-Petterson effect \citep{1975Bardeen}. In addition, the coupling of the X-ray corona to the UV radiation from the disc can propagate other sources of variability \cite[e.g.,][]{Papadakis2001,Arevalo2006}.
These models cannot be tested in detail with the DR1 survey data and require longer pointed observations. Future studies, including all the eROSITA data, are required to test these models 
\citep{2016Boller, 2022Buchner}.
The black hole mass accretion rate is partially converted to radiation with the radiative efficiency $\eta$. A variable emission region of mass M is limited to releasing energy $\rm \Delta L \Delta t < M c^2$. Considering the light travel time and optical depth, gives \citep{CavalloRees1978,1979Fabian}: $$\Delta L / \Delta t<1.6 \times 10^{41} (\eta/0.1)\, \mathrm{erg\,s^{-2}}.$$
As a pilot study, we search for the presence of such strong rapid flux changes in the two most variable AGN, considering their redshift. They do not violate this limit.

The most variable AGN in our sample is PKS 0558$-$504 with AMP of $\rm 3.45~\sigma$ and NEV of $\rm 1.97~\sigma$. The source shows a drop in count rate by a factor of about three within about 
15,000 seconds. PKS 0447$-$439 shows variations by a factor of up to 2 during the eRASS1 observations. 
The AMP and NEV are $\rm 2.18~\sigma\ and\ 2.21~\sigma$, respectively (see Fig.\,\ref{fig:AGN_DR1_e1-e5}).

\begin{figure*}
        \includegraphics[width=9.3cm]{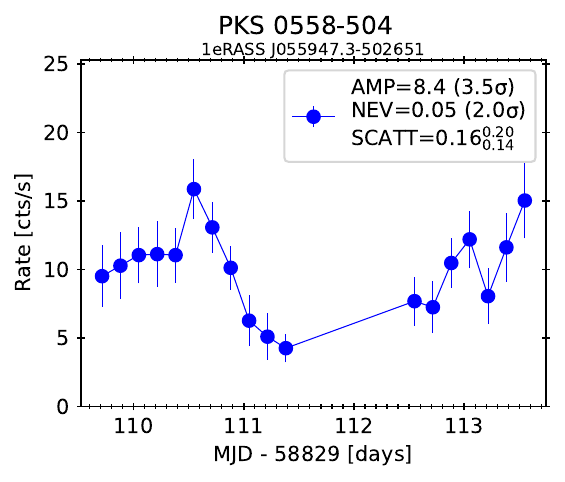}
        \includegraphics[width=9.3cm]{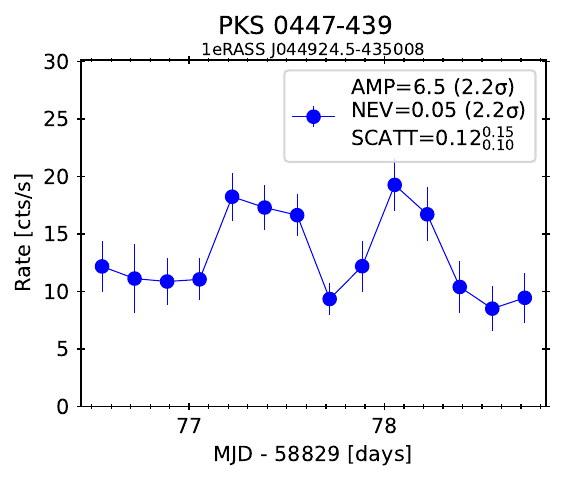}
        \caption{Light curve examples for DR1 sources classified as AGN. The left panel shows the light curve of PKS 0558$-$504, and the right panel is for PKS 0447$-$439
        (c.f. Sect.~\ref{sec:agns}).
               }
        \label{fig:AGN_DR1_e1-e5}
\end{figure*}

\section{Summary}\label{sec:summary}

We selected 128,669 sources with at least tencounts out of DR1, after excluding an area of about 226 square degrees centred on the SEP. We systematically searched for variability in the light curves of the sources using three types of tests: bexvar, AMP and NEV (see Sect.\,\ref{sec:assessing}). According to bexvar 1342, sources are variable, while 808 are considered varying by AMP and 298 by NEV. 
The total number of unique sources is 1709, and 
they form the catalogue of variable sources. The total number of variable sources is less than about 1\% of the complete DR1 catalogue. 

Using more stringent thresholds for AMP and NEV \citepalias{2022Buchner}, only 90 sources demonstrated significant variability via AMP, while 95 did so through NEV. Both subsamples are classified as variable by bexvar, but only 69 sources are in common to all criteria (see Fig.\,\ref{fig:pie_diagram}).
We have performed cross-matches with existing X-ray catalogues. We have identified 258, 318, 598, and 120 sources within 15\arcsec\ have a counterpart in  4XMM DR13, 2SXPS, \ROSAT/2RXS, and CSC2.1 catalogues. However, a more detailed analysis is required to asses whether these are the same sources or not. Only 27 sources are common to all of these catalogues, and 882 sources (52\%) of the eROSITA DR1 variability catalogue  are new X-ray detections.

Using the catalogues of multi-wavelength counterparts of Salvato et. al. (in prep) and \citet{dr1hamstar}, we classified the sources using existing catalogues of LMXB, HMXB, QSOs, etc. About 10\% of the sources are extragalactic (AGN and QSO) while 89\% are Galactic, with 70\% classified as coronal stars. We have further classified the stellar component into LMXB, HMXB, and Bright stars and found a match for 18, 11, and 14 sources, respectively.

We have described in more detail the 12 most variable eRASS1 sources by providing light curves and statistical test results, as well as results obtained for these sources with other X-ray observations. We provide additional descriptions for some subclasses in the eRASS1 variability sample. Magnetically active stars are commonly found among the more variable X-ray sources, and often, burst-like flare events are seen from otherwise faint or undetected objects. 
Examples include the star UCAC4 190-003482 that remained inactive throughout the eRASS1 observations except for the last data point, where it exhibited a sudden increase, reaching a count rate of 30\,\cts or the M dwarf G~124$-$44 where a similar event was detected.
As an example of X-ray variability for X-ray binaries, we describe the case of SMC\,X$-$1 as detected during the eRASS1 observations. Finally, we also provide notes on two AGNs in the eRASS1 variability sample: PKS 0558$-$504 and PKS 0447$-$439.

In this paper, we have analysed the variability of eRASS1 sources on a timescale of a few days only. To study the physics of variable sources, we need more deeply pointed observations with other X-ray missions or at least the final depth of the eRASS:8 observations. The timescale of the eRASS1 observations is not representative of the timescales of the expected upcoming eRASS catalogues. 
The DR1 variability catalogue is an excellent resource for follow-up observations with telescopes such as \xmm, \chandra, or \swift.

\section{Data availability}\label{sec:data} 

The full catalogue will be available in electronic form at the CDS.\footnote{ftp to cdsarc.u-strasbg.fr (130.79.128.5)} 
The list of columns and their description is presented in the Appendix in  the Table \ref{tab:data-model}.

\begin{acknowledgements}

We are grateful for the referee's detailed report, which greatly enhanced the paper's quality.
This work is based on data from eROSITA, the soft X-ray instrument aboard \SRG, a joint Russian-German science mission supported by the Russian Space Agency (Roskosmos), in the interests of the Russian Academy of Sciences represented by its Space Research Institute (IKI), and the Deutsches Zentrum für Luft- und Raumfahrt (DLR). The \SRG spacecraft was built by Lavochkin Association (NPOL) and its subcontractors, and is operated by NPOL with support from the Max Planck Institute for Extraterrestrial Physics (MPE).
The development and construction of the eROSITA X-ray instrument was led by MPE, with contributions from the Dr. Karl Remeis Observatory Bamberg \& ECAP (FAU Erlangen-Nuernberg), the University of Hamburg Observatory, the Leibniz Institute for Astrophysics Potsdam (AIP), and the Institute for Astronomy and Astrophysics of the University of Tübingen, with the support of DLR and the Max Planck Society. The Argelander Institute for Astronomy of the University of Bonn and the Ludwig Maximilians Universität Munich also participated in the science preparation for eROSITA."
The eROSITA data shown here were processed using the eSASS/NRTA software system developed by the German eROSITA consortium.
MK acknowledges support from DLR FKZ 50OR2307.

\end{acknowledgements}

\bibliographystyle{aa}
%%\bibliography{e1_var}    %%%% this is the bib file
\bibliography{aa49355-24}    %%%% this is the bib file

%%%\bibliography{your_bib_file} 

\onecolumn
\appendix
%\begin{appendix}

\section{Catalogue data model}\label{sec:datamodel}
The catalogue includes the key columns of the original eRASS1 catalogue presented in \citep{Merloni2024}, 
focusing on the 0.2-2.3 keV (medium) band.
It also contains columns that describe the variability of the sources as determined in this paper and a list of columns that are the result of the cross-match with existing catalogues. The ID from these catalogues is reported so that for a user, it will be easy to recover additional information. See Table \ref{tab:data-model} for the description of the columns.

%--------------------------
\begin{table*}[h!]
\tiny
 \caption{eROSITA DR1 Variability catalogue column description.},
  \label{tab:data-model},
\begin{tabular}{ll}
\texttt{IAUNAME}   & String containing the official IAU name of the source\\
\texttt{DETUID}    & String unique detection ID \\
\texttt{SKYTILE}   & Sky tile ID \\
\texttt{ID\_SRC}   & Source ID in each sky tile. Use SKYTILE+ID\_SRC to identify the corresponding source products\\
\texttt{UID}       & Integer unique detection ID. It equals CatID$\times 10^{11}$+SKYTILE$\times 10^5$+ID\_SRC, where CatID is 1 for the 1B detected Main \\
                   & and Supp catalogues and 2 for the 3B detected Hard catalogue.\\
\texttt{UID\_Hard} & Hard catalogue eUID of the source with a strong association \\
                   & or -UID if the association is weak. 0 means no counterpart is found in the Hard catalogue. Only in the 1B Main and Supp. catalogues\\
\texttt{UID\_1B}   & Main or Supp catalogue UID of the source with a strong association, \\
                   & or -UID if the association is weak. 0 means no counterpart is found in the 1B catalogues. Only in the Hard catalogue \\
\texttt{ID\_CLUSTER}  &  Group ID of simultaneously fitted sources\\
\hline
\texttt{RA}           &  Right ascension (ICRS), corrected.\\
\texttt{DEC}          &  Declination (ICRS), corrected. \\
\texttt{RA\_RAW}      &  Right ascension (ICRS), uncorrected.\\
\texttt{DEC\_RAW}     &  Declination (ICRS), uncorrected.\\
\texttt{RA\_LOWERR}   & 1-$\sigma$ lower error of RA.\\
\texttt{RA\_UPERR}    & 1-$\sigma$ upper error of RA.\\
\texttt{DEC\_LOWERR}  & 1-$\sigma$ lower error of DEC.\\
\texttt{DEC\_UPERR}   & 1-$\sigma$ upper error of DEC.\\
\texttt{RADEC\_ERR}   &	Combined positional error, raw output from PSF fitting.\\
\texttt{POS\_ERR}     & 1-$\sigma$ position uncertainty.\\
\texttt{LII}          & Galactic longitude.\\
\texttt{BII}          & Galactic latitude.\\
\texttt{ELON}         & Ecliptic longitude.\\
\texttt{ELAT}         & Ecliptic latitude.\\
\texttt{MJD}          & Modified Julian Date of the observation of the source nearest to the optical axis of the telescope. \\
\texttt{MJD\_MIN}     & Modified Julian Date of the first observation of a source.\\
\texttt{MJD\_MAX}     & Modified Julian Date of the last observation of a source.\\
\hline
\texttt{EXT}                  & Source extent parameter.\\
\texttt{EXT\_ERR}             & 1-$\sigma$ error of EXT.\\
\texttt{EXT\_LOWERR}          & 1-$\sigma$ lower error of EXT.\\
\texttt{EXT\_UPERR}           & 1-$\sigma$ upper error of EXT.\\
\texttt{EXT\_LIKE}            & Extent likelihood.\\
\texttt{DET\_LIKE }           & Detection likelihood.  \\
\texttt{ML\_CTS} 	          & Source net counts. \\
\texttt{ML\_CTS\_ERR} 	      & 1-$\sigma$ combined counts error.  \\
\texttt{ML\_CTS\_LOWERR}      & 1-$\sigma$ lower counts error.  \\
\texttt{ML\_CTS\_UPERR}       & 1-$\sigma$ upper counts error.  \\
\texttt{ML\_RATE}    	      & Source count rate.   \\
\texttt{ML\_RATE\_ERR}	      & 1-$\sigma$ combined count rate error. \\
\texttt{ML\_RATE\_LOWERR}     & 1-$\sigma$ lower count rate error. \\
\texttt{ML\_RATE\_UPERR}      & 1-$\sigma$ upper count rate error.  \\
\texttt{ML\_FLUX}    	      & Source flux.   \\
\texttt{ML\_FLUX\_ERR}	      & 1-$\sigma$ combined flux error.  \\
\texttt{ML\_FLUX\_LOWERR}     & 1-$\sigma$ lower flux error.   \\
\texttt{ML\_FLUX\_UPERR}      & 1-$\sigma$ upper flux error.  \\
\texttt{FLAG\_SP\_SNR}        & Source may lie within an overdense region near a supernova remnant.\\
\texttt{FLAG\_SP\_BPS}        & Source may lie within an overdense region near a bright point source.\\
\texttt{FLAG\_SP\_SCL}        & Source may lie within an overdense region near a stellar cluster.\\
\texttt{FLAG\_SP\_LGA}        & Source may lie within an overdense region near a local large galaxy.\\
\texttt{FLAG\_SP\_GC\_CONS}   & Source may lie within an overdense region near a galaxy cluster.  \\
\texttt{FLAG\_NO\_RADEC\_ERR} & Source contained no \texttt{RA\_DEC\_ERR} in the pre-processed version of the catalogue. \\
\texttt{FLAG\_NO\_CTS\_ERR}   & Source contained no \texttt{ML\_CTS\_0\_ERR} in the pre-processed version of the catalogue. \\
\texttt{FLAG\_NO\_EXT\_ERR}   & Source contained no \texttt{EXT\_ERR} in the pre-processed version of the catalogue. \\
\hline
\texttt{NDATA}     & Number of data points in the light curve. \\
\texttt{AMPL}      & Maximum Amplitude variability (see Sect.\,\ref{subsec:definitions}).\\
\texttt{e\_AMPL}   & Uncertainty on Maximum Amplitude Variability (see Sect.\,\ref{subsec:definitions}).\\
\texttt{s\_AMPL}   & The ratio between AMP and e\_AMP (see Sect.\,\ref{subsec:definitions}).\\
\texttt{NEV}       & Normalised excess variance (see Sect.\,\ref{subsec:definitions}). \\
\texttt{e\_NEV}    & Uncertainty on Normalised excess variance (see Sect.\,\ref{subsec:definitions}). \\
\texttt{s\_NEV}    & The ratio between NEV to e\_NEV (see Sect.\,\ref{subsec:definitions}).\\
\texttt{SCATT} & $\sigma_{bexvar}$  (see Sect.\,\ref{subsec:definitions}).  \\
\texttt{SCATT\_LO} & lower 10\% quantile of $\sigma_{bexvar}$  (see Sect.\,\ref{subsec:definitions}).  \\
\texttt{SCATT\_HI} & higher 10\% quantile of $\sigma_{bexvar}$  (see Sect.\,\ref{subsec:definitions}).  \\
\hline
\end{tabular}
\end{table*}

\twocolumn
\begin{table*}[h!]
\tiny
% \caption{eROSITA DR1 Variability catalogue column description.},
%  \label{tab:data-model},
\begin{tabular}{ll}
\texttt{4XMM\_SourceID}    & Corresponding ID in the 4XMM catalogue.\\
\texttt{4XMM\_RAJ2000}     & Rigth Ascension of the source in the 4XMM catalogue.\\
\texttt{4XMM\_DEJ2000}     & Declination of the source in the 4XMM catalogue.\\
\texttt{Sep\_eRO\_4XMM}    & Separation between the eROSITA source and the corresponding 4XMM source. \\
\texttt{2SXPS\_ID}         & Numerical unique source identifier within 2SXPS.\\
\texttt{2SXPS\_RAJ2000}    & Right Ascension (J2000) of the source in the 2SXPS catalogue.\\
\texttt{2SXPS\_DEJ2000}    & Declination (J2000) of the source in the 2SXPS catalogue.\\
\texttt{Sep\_eRO\_2SXPS}   & Separation between the eROSITA source and the corresponding 2SXPS source.\\
\texttt{2RXS\_SourceID}     & Numerical unique source identifier within \ROSAT/2RXS.\\
\texttt{2RXS\_RAJ2000}     & Right Ascension (J2000) of the source in the \ROSAT/2RXS catalogue.\\
\texttt{2RXS\_DEJ2000}     & Declination (J2000) of the source in the \ROSAT/2RXS catalogue.\\
\texttt{Sep\_eRO\_2RXS}    & Separation between the eROSITA source and the corresponding \ROSAT/2RXS source.\\
\texttt{CSC2p1\_SourceID}     & Numerical unique source identifier within CSC2.1 catalogue.\\
\texttt{CSC2p1\_RAJ2000}   & Right Ascension (J2000) of the source in the CSC2.1 catalogue.\\
\texttt{CSC2p1\_DEJ2000}   & Declination (J2000) of the source in the CSC2.p catalogue.\\
\texttt{Sep\_eRO\_CSC2p1}  & Separation between the eROSITA source and the corresponding CSC2.1 source.\\
\texttt{Hamstar\_SourceID}     & Corresponding Gaia DR3 in the Hamstar catalogue.\\
\texttt{CSC2p1\_RAJ2000}   & Rigth Ascension of the Hamstar source in the Gaia DR3 catalogue.\\
\texttt{CSC2p1\_DEJ2000}   & Declination of the Hamstar source in the Gaia DR3 catalogue.\\
\texttt{Sep\_eRO\_Hamstar}  & Separation between the eROSITA source and the corresponding Gaia DR3 in the Hamstar catalogue.\\
\texttt{GDR3\_SourceID}     & CCorresponding ID from Salvato et al 2025 in Gaia DR3 catalogue.\\
\texttt{GDR3\_RAJ2000}   & Rigth Ascension of the source in the Gaia DR3 catalogue.\\
\texttt{GDR3\_DEJ2000}   & Declination of the  source in the Gaia DR3 catalogue.\\
\texttt{Sep\_eRO\_GDR3}  & Separation between the eROSITA source and the corresponding Gaia DR3 in the Salvato et al 2025 catalogue.\\
\texttt{BSC5\_SourceID}     & CCorresponding ID from Salvato et al 2025 in BSC5 catalogue.\\
\texttt{BSC5\_RAJ2000}   & Rigth Ascension of the source in the Gaia DR3 catalogue.\\
\texttt{BSC5\_DEJ2000}   & Declination of the  source in the Gaia DR3 catalogue.\\
\texttt{Sep\_GDR3\_BSC5}  & Separation between the Gaia GDR3 counterpart of the eROSITA source and the corresponding BSC5 source.\\
\texttt{VV\_SourceID}     & CCorresponding ID from Salvato et al 2025 in the Veron \& Veron catalogue.\\
\texttt{VV\_RAJ2000}   & Rigth Ascension of the source in the Veron \& Veron catalogue.\\
\texttt{VV\_DEJ2000}   & Declination of the  source in the Veron \& Veron catalogue.\\
\texttt{Sep\_GDR3\_VV}  & Separation between the Gaia GDR3 counterpart of the eROSITA source and the corresponding Veron \& Veron source.\\
\texttt{LMXB\_SourceID}     & CCorresponding ID from Salvato et al 2025 in the LMXB catalogue.\\
\texttt{LMXB\_RAJ2000}   & Right Ascension of the source in the LMXB catalogue.\\
\texttt{LMXB\_DEJ2000}   & Declination of the  source in the LMXB catalogue.\\
\texttt{Sep\_GDR3\_LMXB}  & Separation between the Gaia GDR3 counterpart of the eROSITA source and the corresponding LMXB source.\\
\texttt{HMXB\_SourceID}     & CCorresponding ID from Salvato et al 2025 in the HMXB catalogue.\\
\texttt{HMXB\_RAJ2000}   & Right Ascension of the source in the LMXB catalogue.\\
\texttt{HMXB\_DEJ2000}   & Declination of the  source in the HMXB catalogue.\\
\texttt{Sep\_GDR3\_HMXB}  & Separation between the Gaia GDR3 counterpart of the eROSITA source and the corresponding HMXB source.\\

\end{tabular}
\end{table*}

\end{document}